\documentstyle[epsf,seceqn]{elsart}
\journal{Nuclear Physics B}  
\newcommand{\com}[1]{ \par }
  
\newcommand{\msb}{$\overline{\mathrm{MS}}$ }
\newcommand{\mrm}[1]{\mbox{\rm #1}}

\def\cslice{Giele:1992vf,Giele:1993dj,Aversa:1990uv,Kramer:1989mc,Baer:1989jg,Fabricius:1981sx,Gutbrod:1984qa}
\def\cmbrun{Santamaria:1997kd,Bilenkii:1995ad,Rodrigo:1996gw,MartiiGracia:1997ak,Rodrigo:1997ha,Rodrigo:1997gv,Rodrigo:1997gy,Fuster:1997ah,Abreu:1997ey}
\def\caachen{Bernreuther:1997jn,Brandenburg:1997pu}
\def\cnason{Oleari:1997az,Nason:1997nu,Nason:1997nw,Nason:1997tz}
\def\ccambridge{Dokshitser:1997in,Bentvelsen:1998ug,Moretti:1998qx}
\def\courcambridge{Rodrigo:1998nk,Rodrigo:1998vq,Bilenky:1998dw,Bilenky:1998dx}
\def\cdelphicambridge{Cabrera:1998xx}
\def\chiggs{Chetyrkin:1996pd,Chetyrkin:1997iv,Chetyrkin:1997vj}
\newcommand{\mafigura}[4]{
  \begin{figure}[hbtp]
    \begin{center}
      \epsfxsize=#1 \leavevmode \epsffile{#2}
    \end{center}
    \caption{#3}
\label{#4}
  \end{figure} }
\def\non{\nonumber}
\newcommand{\beq}{\begin{equation}}
\newcommand{\eeq}{\end{equation}}
\newcommand{\bea}{\begin{eqnarray}}
\newcommand{\eea}{\end{eqnarray}}
\newcommand{\bes}{\begin{eqnarray*}}
\newcommand{\ees}{\end{eqnarray*}}   
\newcommand{\eq}[1]{eq.~(\ref{#1})}
\newcommand{\as}{\alpha_s}
\newcommand{\api}{\frac{\alpha_s}{\pi}}
\newcommand{\yc}{y_{c}}
\newcommand{\rb}{r_b}
\newcommand{\el}[1]{^{(#1)}}
\newcommand{\ql}{\ell}
\begin{document}
\begin{frontmatter}
{\flushright FTUV/99-2\\
 \flushright IFIC/99-2\\[0.5cm]}
\title{Quark-mass effects for jet production in
$e^+ e^-$ collisions at the next-to-leading order: results and applications}
\author[Firenze]{German Rodrigo\thanksref{th:Valencia}},
\author[Prague]{Mikhail Bilenky\thanksref{Dubna}} and
\author[Valencia]{Arcadi Santamaria}
\address[Firenze]{INFN - Sezione di Firenze, Largo E. Fermi 2,
I-50125 Firenze, Italy}
\address[Prague]{Institute of Physics, AS CR, 18040 Prague 8, and
Nuclear Physics Institute,
AS CR, 25068 \v{R}e\v{z}(Prague), Czech Republic}
\address[Valencia]{Departament de F\'{\i}sica Te\`orica, IFIC,
CSIC-Universitat de Val\`encia,\\ E-46100 Burjassot, Val\`encia, Spain}
\thanks[th:Valencia]{On leave from Departament de F\'{\i}sica Te\`orica,
Universitat de Val\`encia, Val\`encia, Spain}
\thanks[Dubna]{On leave from Laboratory of Nuclear Problems, JINR, 141980
Dubna, Russian Federation}

\begin{abstract}
We present a detailed description of our calculation of
next-to-leading order QCD corrections
to heavy quark production in $e^+ e^-$ collisions including mass effects.
In particular, we study the observables $R_3^{b\ql}$ and $D_2^{b\ql}$ in the
E, EM, JADE and DURHAM jet-clustering algorithms and show how one
can use these observables to obtain $m_b(m_Z)$ from data at the $Z$ peak. 
\end{abstract}

\end{frontmatter}

\noindent
PACS number(s): 12.15.Ff,12.38.Bx,13.87.Ce,14.65.Fy

\section{Introduction}

The origin of fermion masses is one of the unresolved
puzzles in present high energy physics. To be able to answer this question 
one needs to know the values of the quark masses with high accuracy. However,
due to confinement, quarks do not appear as free particles in nature,
and therefore, the definition of their mass is ambiguous. Quark masses
can be understood more easily if they are interpreted like coupling constants 
rather than fixed inertial parameters and, therefore,  quark 
masses can run if measured at different scales. Moreover, in the standard 
model (SM) all
fermion masses come from Yukawa couplings and those also run with
the energy. This has very important phenomenological consequences in
Higgs boson searches since the parameter that governs the decay rate of
the Higgs boson to bottom quarks~\cite{\chiggs} is the running mass of the 
$b$-quark at the
$m_H$ scale, $m_b(m_H)$, which is much smaller than the quark mass extracted 
at threshold. On the other hand, from
the model building point of view, to test fermion mass models one has to run 
masses  extracted at quite different scales up to the same scale and compare 
them with the same ``ruler''~\cite{Rodrigo:1995az}. This way, for 
instance, one can check that, in some grand unified models, although 
the $b$-quark mass and the tau mass are very different at threshold energies, 
they could be equal at the unification
scale~\cite{Langacker:1994xb,Santamaria:1997kd}. 

However, although predicted by quantum field theory, the
possibility of testing experimentally the running of fermion masses has not 
been considered until very recently~\cite{\cmbrun}.
The reason being that for energies much higher than the fermion 
mass the mass effects become negligible for most observables. 
In principle, it is clear that the running of quark masses will be checked,
once the Higgs is discovered, by measuring directly the Yukawa couplings in
Higgs decays. We think it is also interesting to explore the
possibility of measuring quark masses far away from threshold in order to
check the running. 

In \cite{Bilenkii:1995ad,Chrin.moriond:1993,Fuster.jaca:1994,Abreu:1993mc,Buskulic:1995wb}
it was shown that mass effects in three-jet production at the $Z$ peak are 
large enough to be measured but in~\cite{Bilenkii:1995ad} we also showed that 
a next-to-leading order (NLO) 
calculation of three-jet ratios including mass effects was necessary in order 
to extract a meaningful value for the $b$-quark mass. This is because 
the leading order (LO) calculation does not distinguish among the different 
quark mass definitions (e.g. perturbative pole mass or \msb running mass).

Although jet production of heavy quarks has been considered in a large 
variety of processes, there are very few NLO calculations
of jet production taking into account complete mass effects.
In~\cite{Beenakker:1989bq} those were calculated for gluon-gluon fusion
in proton-antiproton collisions,
while in~\cite{Laenen:1993zk} the complete NLO corrections were
computed for virtual-photon production of heavy quarks in deep-inelastic
electron-proton scattering. 
However, until very recently, no full calculation 
of jet production of heavy quarks in $e^+ e^-$ collisions was available at 
NLO\footnote{Completely inclusive quantities as the total cross-section were
fully known at order 
$\as$~\cite{Shifman:1978zq,Reinders:1981sy,Chang:1982qq,Laermann:1980qk}
and some leading quark mass effects in those quantities
were also known up to order $\as^3$~\cite{Chetyrkin:1990kr,Chetyrkin:1993tx}.
Quark mass effects in three-jet final states, however, were only known at 
leading
order~\cite{Laermann:1980qk,Ioffe:1978dc,Kramer:1980pg,Rizzo:1980cc,Nilles:1980ic,Arbuzov:1992pr,Grunberg:1980ru,Ballestrero:1992ed,Ballestrero:1994dv,Olsen:1997sk}.
}.
In \cite{Rodrigo:1996gw,Rodrigo:1997gy} we have presented the final results of
such a calculation and have shown
how can one use the DURHAM ($k_T$) clustering algorithm to extract the 
$b$-quark mass at the
$m_Z$ scale. Finally in~\cite{Abreu:1997ey} the DELPHI collaboration
has presented the measurement of $m_b(m_Z)$. The obtained value is in
good agreement with low energy measurements and running of the mass from
$\mu=m_b$ to $\mu=m_Z$ as predicted by QCD with 5 flavours.
Our NLO calculation was also used in~\cite{Abreu:1997ey} to check the
universality of the strong interactions.
In the meanwhile two more groups,
ref.~\cite{\caachen} and ref.~\cite{\cnason},
have also presented NLO results of jet production of heavy quarks in
$e^+ e^-$ collisions by using different calculational schemes. On the
other hand also the different experimental collaborations have started
to use NLO calculations of mass effects in order to check the universality
of the strong interactions~\cite{Chrisman:1998,Abe:1998kr,Burrows:1998kk}
and have also shown that mass effects are absolutely necessary to explain the 
data. 

In this paper we present a detailed description of our calculation of
the NLO corrections to the rates of jet production of massive quarks in
$e^+ e^-$ collisions for the different jet-clustering algorithms discussed 
in~\cite{Bilenkii:1995ad}. In~\cite{Abreu:1997ey,Rodrigo:1997gy} only the 
DURHAM algorithm was used because it has a very good behaviour from both the 
experimental and the theoretical points of view. It is important to stress 
that not all the observables are equally good to study mass effects. 
In particular observables which suffer from huge NLO corrections cannot
be used to extract any reasonable value of the $b$-quark mass. 

Although, in principle, we will concentrate
on $b$-quark production at the $Z$-peak,
some of the results can also be applied
to $t$-$\bar{t}$ production at the NLC (Next-Linear-Collider). 

In section \ref{sec:lo} we review
the leading order results, give LO amplitudes in $D$-dimensions 
and introduce the jet-clustering algorithms we use. 
In section \ref{sec:nlo} the strategy of the NLO calculation in
the framework of the phase-space slicing 
method~\cite{\cslice}
is explained. Section
\ref{sec:3parton} gives all the contributions coming from three partons,
while in section \ref{sec:4parton} we give all four-parton contributions
and show the cancellation of the infrared (IR)
divergences among three-parton and four-parton contributions.
Finally in section \ref{sec:mb} 
we present numerical results for the finite parts and
review applications of this calculation.
In particular we study two observables, $R_3^{b\ql}$ (the ratio of the 
three-jet rate containing the $b$-quark to the three-jet rate of light quarks) 
and $D_2^{b\ql}$ (the ratio of the differential two-jet rate containing the
$b$-quark to the differential rate of light quarks), computed in the different 
jet-clustering algorithms and show how they can be used to extract 
$m_b(m_Z)$ from data at the $Z$-peak.
In the appendices we 
collect some of the formulae used in the main calculation. Thus, 
in appendix \ref{ap:loop}
we give the most relevant one-loop scalar functions that appear in the
calculation of the virtual corrections. In appendix \ref{ap:ps} we review
the calculation of phase-space in D-dimensions. Finally in appendix
\ref{ap:divs} we collect some of the IR divergent integrals which appear
in the calculation of four-parton contributions.

\section{The decay $ Z \rightarrow 3 jets$ with heavy quarks at the 
leading order}
\label{sec:lo}

\subsection{Decay into two and three jets 
and jet-clustering algorithms}

The lowest order contribution to the $Z$-boson decay into three jets
in the parton picture is given by the decay 
$Z \rightarrow b \bar{b} g$. This partonic process has an infrared
singularity because a massless gluon could be radiated with zero 
energy\footnote{In the limit of massless
quarks there are collinear singularities as well. 
In the massive case the quark-gluon collinear
singularities are softened into logarithms of the quark mass.
Gluon-gluon collinear singularities appear in our calculations at
the NLO and will be discussed in the following sections.}.
It is well known~\cite{Bloch:1937pw,Kinoshita:1962ur,Lee:1964is}, 
however, that this divergence is exactly
canceled by the IR-divergent part of the two-parton decay width of the
$Z$-boson, $Z \rightarrow b \bar{b}$, at the order ${\cal{O}}(\as)$.
In the last process the IR divergence occurs 
due to massless gluons running in the loops.

Thus, in order to define IR-finite observables,
we need to introduce some {\it resolution parameter}, $\yc$,
and to split the full three-parton 
phase-space into the two-jet region, containing
soft gluon emission, and the remaining three-jet region.
The sum of the three-parton decay probability integrated over
the two-jet part of the phase-space and the two-parton decay width
at the order ${\cal{O}}(\as)$ defines the IR finite 
{\it two-jet decay width}, $\Gamma^b_{2j}$, at the next-to-leading order.
The integration over the rest of the three-parton 
phase-space gives the {\it three-jet decay width},
$\Gamma^b_{3j}$, at the leading order and it is also IR finite.

Although both IR-finite $\Gamma^b_{2j}(\yc)$ and 
$\Gamma^b_{3j}(\yc)$ depend on the resolution parameter, it is obvious
that the sum of the two-jet and three-jet decay widths is independent
of $\yc$ and is given by the total (inclusive) decay width of the
$Z$-boson into the heavy quarks at the next-to-leading order:
\[
\Gamma^{b}= \Gamma^b_{2j}(\yc)+\Gamma^b_{3j}(\yc)~.
\]
Therefore, at order ${\cal{O}}(\as)$, knowing 
$\Gamma^{b}$ and $\Gamma^b_{3j}(\yc)$ we can obtain
$\Gamma^b_{2j}(\yc)$ as well.
The analytical expression for the $\Gamma^{b}$
with quark-mass corrections can be found 
in~\cite{Bilenkii:1995ad,Chetyrkin:1990kr,Chetyrkin:1993tx,Jersak:1982sp,Djouadi:1990uk,Chetyrkin:1996ia}.

In the same way, at the next-to-next-to-leading order we can write for the
total width:
\begin{equation}
\label{eq:twojets}
\Gamma^{b}= \Gamma^b_{2j}(\yc)+\Gamma^b_{3j}(\yc)+\Gamma^b_{4j}(\yc)~.
\end{equation}
Now the four-jet decay width, $\Gamma^b_{4j}(\yc)$, receives 
contributions from four-parton processes. The three-jet 
decay width, $\Gamma^b_{3j}(\yc)$, gets contributions from
order ${\cal O}(\as^2)$ one-loop corrected three-parton processes 
and the soft part of the four-parton processes such that the
sum is IR-finite. Thus, knowing $\Gamma^{b}$, $\Gamma^b_{3j}(\yc)$
and $\Gamma^b_{4j}(\yc)$ we can obtain
$\Gamma^b_{2j}(\yc)$ at order ${\cal O}(\as^2)$.

Let us discuss now the definition of jets in more detail.
The most popular jet definitions used in the analysis of
the $e^+e^-$-annihilation during the last years 
are based on the so-called {\it jet-clustering algorithms}.
These algorithms have to be applied to define jets 
in both, the theoretical calculations at the parton
level, and in the analysis of the 
bunch of real particles observed at experiment.

In the jet-clustering algorithms jets are usually defined as follows:
starting from a bunch of $n$ particles\footnote{In 
what follows we will use the word
``particles'' for both partons and real particles.}
with four-momenta $p_i,~(i=1..n)$ one computes a resolution
function depending on the momenta of two particles, 
for example,
\[
y_{ij}=2 \frac{E_i E_j}{s} (1-\cos \theta_{ij})~,
\]
for all pairs $(i,~j)$ of particles. Then one takes the minimum of
all $y_{ij}$ and, if it satisfies that it is smaller than a given quantity 
$\yc$ (the resolution parameter, y-cut), the two particles 
which define the minimal $y_{ij}$ are regarded
as belonging to the same 
jet\footnote{The assignment of particles to the jet in the recently proposed
Cambridge scheme is more involved, see
refs~\cite{\ccambridge}
for details.}. Then they are recombined into a new
pseudo-particle with the four-momentum defined according
to some rule, for example,
\[
p_k = p_i + p_j~.
\]
After this first step one has a bunch of $n-1$ (pseudo)particles and
the algorithm is applied again and again until all the remaining
(pseudo)particles satisfy the criteria $y_{ij} > \yc$. The final number 
of (pseudo)particles is the number of jets in the considered event and 
the momenta of the (pseudo)particles define the event kinematics.

In theoretical calculations one can define a
jet cross-section or a decay width into jets as a function of $\yc$,
which are computed at the parton level, by following exactly
the algorithm described above. Jet-clustering algorithms
lead automatically to IR finite quantities.
Furthermore, it has been shown in the literature, using Monte Carlo models,
that, for some of the algorithms, the passage from partons to hadrons
(hadronization) does not change much the behaviour of the
jet-observables~\cite{Bethke:1992wk}, thus allowing to compare
theoretical predictions with experimental results. 
We refer to~\cite{Bethke:1992wk,Kunszt:1989km} for a detailed 
discussion and comparison of different successful jet-clustering
algorithms used in $e^+e^-$ annihilation in the case of massless
quarks. Hadronization corrections for the case of massive quarks
have also been studied for the DURHAM algorithm~\cite{Abreu:1997ey}.

In this paper we will present results and
compare them for the four jet-clustering algorithms listed in
table~\ref{tab:jetalg},
where $\sqrt{s}$ is the total center of mass energy.
Already in ref.~\cite{Bilenkii:1995ad} we showed that the $E$-algorithm
has a very peculiar behaviour when massive partons are involved.
It is  because for the same values of particle momenta, the resolution
function, $y_{ij}$, is significantly shifted, comparing with the massless case,
when the quark masses are included.
To avoid these problems in the massive case we introduced in
ref.~\cite{Bilenkii:1995ad} the $EM$-algorithm, which is similar to 
the $E$-algorithm. 
As mentioned above, all observables have a completely different dependence
on $\yc$ for the $E$-algorithm and this can serve as a good test of both 
calculations and data analyses. The application of the new interesting 
Cambridge algorithm~\cite{\ccambridge}
in the massive case is presently under
study~\cite{\courcambridge,\cdelphicambridge}.

\begin{center}
\def\arraystretch{1.5}
\begin{table}
\caption{The jet-clustering algorithms\label{tab:jetalg}}
\begin{tabular}{||l|l|l||}
\hline
Algorithm & Resolution function, $y_{ij}$ & Combination rule\\
\hline\hline
EM & $2(p_i\cdot p_j)/s$ & $p_k= p_i+p_j$ \\
JADE & $2 E_i E_j (1-\cos \theta_{ij})/s$ & $p_k= p_i+p_j$ \\ 
E & $(p_i+p_j)^2/s$ & $p_k = p_i+p_j$ \\
DURHAM  & $2 \min(E_i^2,E_j^2)(1-\cos \theta_{ij})/s\ \ $ & 
$p_k = p_i+p_j$\\
\hline
\end{tabular}
\end{table}
\end{center}    

It is convenient to parameterize the final IR-finite
result for the three-jet decay width of
the $Z$-boson in the following general form ($D=4$):
\beq
\label{eq:gamma3jets}
\Gamma^b_{3j} = \sqrt{s} \frac{\alpha}{64 s_W^2 c_W^2 } \api
\left[ g_V^2(b) H_V(\yc,\rb) + g_A^2(b) H_A(\yc,\rb)\right]~,
\eeq
where the functions $H_{V(A)}(\yc,\rb)$
depend, obviously, on what jet-clustering algorithm has been used
to define the three-jet region of phase-space.
We have introduced the notation $\rb = M_b^2/s$, where $s=m_Z^2$
for the $Z$-decay.
Here and in the following $M_b$ will stand for  the perturbative pole mass
of the $b$-quark, while the running mass will be denoted as $m_b(\mu)$.
The vector and axial-vector neutral-current couplings $g_V(f)$
and $g_A(f)$ of a fermion, $f$, in the Standard Model are given by
\bea
g_V(f)=2 T^3_f - 4 Q_f s^2_W,\qquad g_A(f)=-2 T^3_f,
\label{eq:gvga}
\eea
being $s_W~(c_W)$ the sine (cosine) of the weak mixing angle,
$T^3_f$ the third component of the weak isospin and $Q_f$ the 
electric charge of the fermion.

We can expand the $H_{V(A)}$ functions in the strong coupling constant,
$\as$  
\beq
H_{V(A)}(\yc,\rb) =  H^{(0)}_{V(A)}(\yc,\rb) +
\api H^{(1)}_{V(A)}(\yc,\rb)+\cdots~, 
\label{eq:hvaexp}
\eeq
where $H^{(0)}_{V(A)}$ and  $H^{(1)}_{V(A)}$ are the LO and NLO 
contributions, respectively.
Furthermore, in order to see more clearly the size of quark-mass
effects, when the masses are
small with respect to the center of mass energy, we found it
convenient to rewrite these functions factoring the dominant 
$r_b$-dependence
\beq
H^{(i)}_{V(A)}(\yc,\rb) \equiv 
A^{(i)}_{V(A)}(\yc)+{\rb}B^{(i)}_{V(A)}(\yc,\rb)~.
\label{eq:Bform}
\eeq
Note that \eq{eq:Bform} is not an expansion in $r_b$ because
the functions $B^{(i)}_{V(A)}(\yc,\rb)$ contain the exact residual 
$r_b$-dependence.

In the limit of zero quark masses, $\rb=0$, chirality is conserved
and the two lowest-order functions $H\el{0}_V(\yc,\rb)$ 
and $H\el{0}_A(\yc,\rb)$ become identical
\beq
H^{(0)}_V(\yc,0) =  H^{(0)}_A(\yc,0) \equiv A^{(0)}(\yc)~.
\label{eq:zeromass}
\eeq
This is not true at the NLO where the 
vector and the axial-vector functions,
$A^{(1)}_V$ and $A^{(1)}_A$, are not equal anymore
due to the small contribution of the triangle
diagram~\cite{Hagiwara:1991dx} $V12$ in fig.~\ref{fig:loops}.

The LO and NLO massless contributions, the $A^{(0)}$ and $A^{(1)}_V$
functions\footnote{Note that with our
choice of the normalization $A^{(0)}(\yc) = \frac{1}{2} A(\yc)$
and $A^{(1)}_V(\yc)=\frac{1}{4} B(\yc)$
where $A(\yc)$ and $B(\yc)$ are defined in~\cite{Bethke:1992wk}.},
were calculated in~\cite{Bethke:1992wk} using the results 
from~\cite{Ellis:1981wv}.
In~\cite{Bilenkii:1995ad} the lowest order mass effects, functions 
$B^{(0)}_{V(A)}$, have been calculated analytically for the EM
jet-clustering algorithm and numerically for JADE, DURHAM
and E algorithms.
For completeness, we review in the next section the LO results.
Results for the NLO heavy quark contribution,
in particular the functions $B^{(1)}_{V(A)}$ in~\eq{eq:Bform}
will be presented later.

Concluding this section we would like to make the following remark.
In this paper we discuss the
$Z$-boson decay. In LEP experiments one studies the process
$e^+e^- \rightarrow (Z \gamma^*)\rightarrow b\bar{b} +\cdots$ and, apart
from the resonant 
$Z$-exchange cross-section, there are contributions from the pure
$\gamma$-exchange and from the $\gamma-Z$-interference. 
The non-resonant $\gamma$-exchange contribution 
at the peak is less than 1\% for muon production and in the case
of $b$-quark production there is an additional suppression factor
$Q_b^2=1/9$. In the vicinity of the $Z$-peak the interference
is also suppressed because it is proportional to $Q_b (s-m_Z^2)$.
However, away from the $Z$ resonance
these contributions are important and should be properly taken into account.

The extension of our calculation for 
the cross-section of $e^+e^-$ annihilation into three jets
can be done~\cite{Jersak:1982sp} as follows

\[
\sigma^b_{3j}(s,\yc,\rb)=\api \left[ \sigma^b_V(s) H_V(\yc,\rb)
                             +\sigma^b_A(s) H_A(\yc,\rb) \right]
\label{eq:ee3j}
\]
with the same $H_{V(A)}(\yc,\rb)$ functions introduced in \eq{eq:gamma3jets} 
for the Z-decay. The functions
$\sigma^b_{V(A)}(s)$ have the form ($D=4$):
\bea
\sigma^b_V(s) =  \frac{4\pi\alpha^2}{s} 
  & &\left\{  Q_e^2 Q_b^2 
  +  2 Q_e Q_b \frac{1}{16 s_W^2 c_W^2} g_V(e) g_V(b)Re (\chi (s))
\right.
\non \\
 & & +  \left. \frac{1}{64 s_W^4 c_W^4}g_V(b)^2 
\left[ g_V(e)^2+g_A(e)^2 \right] |\chi(s)|^2 \right\}~,
\non \\ \non \\
\sigma^b_A(s) =  \frac{4\pi\alpha^2}{s} & &\left\{
 \frac{1}{64 s_W^4 c_W^4}g_A(b)^2 
\left[ g_V(e)^2+g_A(e)^2 \right] |\chi(s)|^2 \right\}~,
\label{eq:sigvv}
\eea
where
\[
\chi(s)=\frac{s}{s-m_Z^2+im_Z\Gamma_Z~}.
\]
Also, the numerically important 
QED initial-state radiation (ISR) should be taken into account
in the realistic analysis of the experimental data.
The cross-section of the three-jet production including ISR
can be written as a convolution
\beq
\tilde{\sigma}^b_{3j}(s,\yc,\rb)=
\int \sigma^b_{3j}(s',\yc,\rb) F_{ISR}(s'/s)ds'~,
\eeq
where $F_{ISR}(s'/s)$ is the well-known QED radiator~\cite{Berends:1988ab} 
for the total cross-section and $s'$ denotes the invariant mass of the 
jet final state.

\subsection{$Z \rightarrow 3 jets$ at tree level}
\label{sec:3jets0}
In the NLO calculation of the $Z$-decay width into three-jets,
which will be presented in the next sections,
we encounter ultraviolet (UV) and infrared singularities.
We use dimensional regularization to regularize both types
of divergences and therefore most of the
calculation has to be done in arbitrary $D$-dimensions.
The calculation of $\Gamma^b_{3j}(\yc)$ at order $\as$
is a pure tree-level calculation
and it does not have any IR problem, since the soft gluon region 
is excluded from the three-parton phase-space. 
Therefore, the calculation can be safely done in four dimensions. 
However, the later use of the LO results in some steps of the
NLO calculation require the tree-level results in $D$-dimensions
that we present in the following.

\mafigura{7.cm}{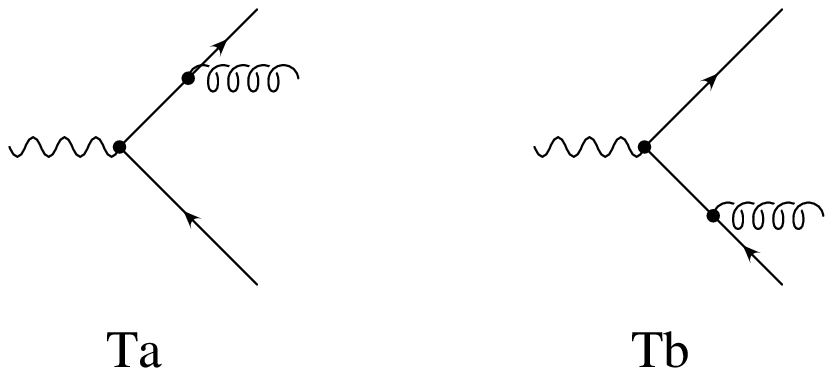}
{Tree-level diagrams contributing to $Z\rightarrow b\bar{b}g$.}
{fig:zbbg}

In $D=4-2\epsilon$ dimensions
the Born transition probability for the three-parton process, 
$Z \rightarrow b\ \bar{b}\ g$, summed over final colours
and spins and averaged over initial spins is equal to
\begin{equation}
\label{eq:m2bbg}
\overline{\sum} \mid M_{b\bar{b}g} \mid^2
= \frac{g^2}{16 c_W^2} g_s^2 \frac{C_F N_C}{3-2\epsilon} \: T_{b\bar{b}g}~,
\end{equation}
where $g$ the $SU(2)$ coupling constant, $g_s$ is the strong coupling
constant, $C_F=4/3$ and $N_C=3$ are $SU(3)$ group invariant factors and 
\bea
T_{b\bar{b}g}(y_{13},y_{23}) &=& 16 \left\{ \frac{h_p}
{y_{13}^2 y_{23}^2} T_{b\bar{b}} 
+ g_V^2(b) (1-\epsilon) \left[\frac{1}{2} (1-\epsilon) 
\left( \frac{y_{13}}{y_{23}} + \frac{y_{23}}{y_{13}} \right)
- \epsilon \right] \right. \non \\
& & \left. \hskip -1.5cm + g_A^2(b) (1-\epsilon) \left[ 
\frac{1}{2} (1-\epsilon+2r_b)
\left( \frac{y_{13}}{y_{23}} + \frac{y_{23}}{y_{13}} \right)
+ 2r_b-\epsilon \right] \right\}~,
\label{eq:t1}
\eea
with
\[
y_{13}=\frac{2(p_1\cdot p_3)}{s}~, \qquad
y_{23}=\frac{2(p_2\cdot p_3)}{s}~,
\]
where $p_1$, $p_2$ and $p_3$ are the four-momenta of the quark, 
the antiquark and the gluon, respectively.
The function $h_p(y_{13},y_{23})$ 
defines the three-body phase-space boundaries (See \eq{eq:bornPS})
and the dimensionless function 
\beq
T_{b\bar{b}} = g_V^2(b)(1-\epsilon+2r_b)
+g_A^2(b)(1-\epsilon)(1-4r_b)~,
\label{eq:t0}
\eeq
is related to the Born two-parton process, $Z \rightarrow b\ \bar{b}$,
transition probability through
\beq
\overline{\sum} \mid M_{b\bar{b}} \mid^2
= \frac{g^2}{16 c_W^2} \frac{N_C}{3-2\epsilon} \: 4 s \: T_{b\bar{b}}~.
\eeq
Averaging over the $Z$-boson polarizations we took into account that
in arbitrary space-time dimensions the number of spin degrees of freedom is 
$D-1=3-2\epsilon$.
 \begin{table}[htb]
  \caption{The leading order functions 
  (see eqs.~(\ref{eq:gamma3jets}, \ref{eq:hvaexp}, \ref{eq:Bform})) for
  the E algorithm.\label{tab:e0}}
 \begin{tabular}{|c|r|r|r|r|r|r|}
 \hline
 $y_c$   & $A^{(0)}\: \:$ &
 \multicolumn{3}{c|}{$B^{(0)}_V/A^{(0)}$}
 \\\cline{3-5}&
  & $M_b=3$GeV & $M_b=4$GeV & $M_b=5$GeV
 \\ \hline
  0.01 &   18.627 &   37.811 &   39.353 &   42.432  \\
  0.02 &   12.209 &   22.195 &   21.645 &   21.723 \\
  0.03 &    9.069 &   16.917 &   16.314 &   15.840 \\
  0.04 &    7.118 &   13.540 &   13.560 &   13.135 \\
  0.05 &    5.765 &   11.390 &   11.466 &   11.452 \\
  0.06 &    4.764 &    9.932 &    9.981 &   10.043 \\
  0.07 &    3.990 &    8.889 &    8.920 &    8.960 \\
  0.08 &    3.374 &    8.118 &    8.138 &    8.162 \\
  0.09 &    2.873 &    7.540 &    7.551 &    7.565 \\
  0.10 &    2.458 &    7.108 &    7.113 &    7.118 \\
 \hline
 $y_c$   & $A^{(0)} \: \:$ &
 \multicolumn{3}{c|}{$B^{(0)}_A/A^{(0)}$}
 \\\cline{3-5}&
 & $M_b=3$GeV & $M_b=4$GeV & $M_b=5$GeV
 \\ \hline
  0.01 &   18.627 &   32.506 &   33.827 &   36.578 \\
  0.02 &   12.209 &   17.188 &   16.532 &   16.462 \\
  0.03 &    9.069 &   12.084 &   11.414 &   10.847 \\
  0.04 &    7.118 &    8.838 &    8.812 &    8.321 \\
  0.05 &    5.765 &    6.797 &    6.838 &    6.777 \\
  0.06 &    4.764 &    5.436 &    5.455 &    5.481 \\
  0.07 &    3.990 &    4.478 &    4.486 &    4.495 \\
  0.08 &    3.374 &    3.785 &    3.785 &    3.785 \\
  0.09 &    2.873 &    3.280 &    3.274 &    3.267 \\
  0.10 &    2.458 &    2.914 &    2.905 &    2.893 \\
 \hline
 \end{tabular}
 \end{table} 

 \begin{table}[htb]
  \caption{The leading order functions 
  (see eqs.~(\ref{eq:gamma3jets}, \ref{eq:hvaexp}, \ref{eq:Bform})) for
  the EM algorithm.\label{tab:em0}}
 \begin{tabular}{|c|r|r|r|r|r|r|}
 \hline
 $y_c$   & $A^{(0)}\: \:$ &
 \multicolumn{3}{c|}{$B^{(0)}_V/A^{(0)}$}
 \\\cline{3-5}&
  & $M_b=3$GeV & $M_b=4$GeV & $M_b=5$GeV
 \\ \hline
  0.01 &   18.628 &  -18.708 &  -18.778 &  -18.444 \\
  0.02 &   12.209 &  -12.406 &  -13.178 &  -13.529 \\
  0.03 &    9.069 &  -10.081 &  -10.637 &  -11.167 \\
  0.04 &    7.118 &   -9.464 &   -9.480 &   -9.813 \\
  0.05 &    5.765 &   -9.162 &   -9.161 &   -9.168 \\
  0.06 &    4.764 &   -9.018 &   -9.016 &   -9.013 \\
  0.07 &    3.990 &   -8.973 &   -8.970 &   -8.966 \\
  0.08 &    3.374 &   -8.997 &   -8.993 &   -8.988 \\
  0.09 &    2.873 &   -9.073 &   -9.068 &   -9.062 \\
  0.10 &    2.458 &   -9.192 &   -9.186 &   -9.178 \\
 \hline
 $y_c$   & $A^{(0)} \: \:$ &
 \multicolumn{3}{c|}{$B^{(0)}_A/A^{(0)}$}
 \\\cline{3-5}&
  & $M_b=3$GeV & $M_b=4$GeV & $M_b=5$GeV
 \\ \hline
  0.01 &   18.628 &  -23.675 &  -23.678 &  -23.261 \\
  0.02 &   12.209 &  -17.214 &  -17.931 &  -18.215 \\
  0.03 &    9.069 &  -14.764 &  -15.269 &  -15.736 \\
  0.04 &    7.118 &  -14.042 &  -14.008 &  -14.279 \\
  0.05 &    5.765 &  -13.648 &  -13.598 &  -13.544 \\
  0.06 &    4.764 &  -13.419 &  -13.370 &  -13.308 \\
  0.07 &    3.990 &  -13.295 &  -13.247 &  -13.184 \\
  0.08 &    3.374 &  -13.246 &  -13.198 &  -13.135 \\
  0.09 &    2.873 &  -13.255 &  -13.205 &  -13.142 \\
  0.10 &    2.458 &  -13.310 &  -13.260 &  -13.195 \\
 \hline
 \end{tabular}
 \end{table} 

 \begin{table}[htb]
  \caption{The leading order functions 
  (see eqs.~(\ref{eq:gamma3jets}, \ref{eq:hvaexp}, \ref{eq:Bform})) for
  the JADE algorithm.\label{tab:jade0}}
 \begin{tabular}{|c|r|r|r|r|r|r|}
 \hline
 $y_c$   & $A^{(0)}\: \:$ &
 \multicolumn{3}{c|}{$B^{(0)}_V/A^{(0)}$}
 \\\cline{3-5}&
  & $M_b=3$GeV & $M_b=4$GeV & $M_b=5$GeV
 \\ \hline
  0.01 &   18.627 &  -34.488 &  -31.314 &  -28.526 \\
  0.02 &   12.209 &  -25.045 &  -24.050 &  -22.805 \\
  0.03 &    9.069 &  -20.158 &  -19.968 &  -19.527 \\
  0.04 &    7.118 &  -17.410 &  -17.308 &  -17.191 \\
  0.05 &    5.765 &  -15.675 &  -15.596 &  -15.501 \\
  0.06 &    4.764 &  -14.491 &  -14.437 &  -14.384 \\
  0.07 &    3.990 &  -13.613 &  -13.577 &  -13.529 \\
  0.08 &    3.374 &  -12.924 &  -12.902 &  -12.870 \\
  0.09 &    2.873 &  -12.393 &  -12.373 &  -12.346 \\
  0.10 &    2.458 &  -11.979 &  -11.964 &  -11.943 \\
 \hline
 $y_c$   & $A^{(0)}\: \:$ &
 \multicolumn{3}{c|}{$B^{(0)}_A/A^{(0)}$}
 \\\cline{3-5}&
  & $M_b=3$GeV & $M_b=4$GeV & $M_b=5$GeV
 \\ \hline
  0.01 &   18.627 &  -39.412 &  -36.144 &  -33.244 \\
  0.02 &   12.209 &  -29.823 &  -28.753 &  -27.415 \\
  0.03 &    9.069 &  -24.817 &  -24.560 &  -24.035 \\
  0.04 &    7.118 &  -21.968 &  -21.802 &  -21.606 \\
  0.05 &    5.765 &  -20.142 &  -20.002 &  -19.832 \\
  0.06 &    4.764 &  -18.876 &  -18.764 &  -18.638 \\
  0.07 &    3.990 &  -17.921 &  -17.829 &  -17.711 \\
  0.08 &    3.374 &  -17.161 &  -17.086 &  -16.986 \\
  0.09 &    2.873 &  -16.564 &  -16.492 &  -16.399 \\
  0.10 &    2.458 &  -16.088 &  -16.022 &  -15.937 \\
  \hline
 \end{tabular}
 \end{table}

 \begin{table}[htb]
  \caption{The leading order functions 
  (see eqs.~(\ref{eq:gamma3jets}, \ref{eq:hvaexp}, \ref{eq:Bform})) for
  the DURHAM algorithm.\label{tab:durham0}}
 \begin{tabular}{|c|r|r|r|r|r|r|}
 \hline
 $y_c$   & $A^{(0)}\: \:$ &
 \multicolumn{3}{c|}{$B^{(0)}_V/A^{(0)}$}
 \\\cline{3-5}&
  & $M_b=3$GeV & $M_b=4$GeV & $M_b=5$GeV
 \\ \hline
  0.01 &    7.836 &  -27.862 &  -27.092 &  -26.202 \\
  0.02 &    5.127 &  -20.038 &  -19.800 &  -19.509 \\
  0.03 &    3.815 &  -16.760 &  -16.652 &  -16.517 \\
  0.04 &    3.003 &  -14.889 &  -14.833 &  -14.762 \\
  0.05 &    2.440 &  -13.659 &  -13.628 &  -13.589 \\
  0.06 &    2.023 &  -12.780 &  -12.764 &  -12.743 \\
  0.07 &    1.701 &  -12.117 &  -12.110 &  -12.101 \\
  0.08 &    1.443 &  -11.598 &  -11.597 &  -11.595 \\
  0.09 &    1.233 &  -11.180 &  -11.183 &  -11.186 \\
  0.10 &    1.058 &  -10.836 &  -10.842 &  -10.849 \\
 \hline
 $y_c$   & $A^{(0)} \: \:$ &
 \multicolumn{3}{c|}{$B^{(0)}_A/A^{(0)}$}
 \\\cline{3-5}&
  & $M_b=3$GeV & $M_b=4$GeV & $M_b=5$GeV
 \\ \hline
  0.01 &    7.836 &  -32.216 &  -31.337 &  -30.316 \\
  0.02 &    5.127 &  -24.256 &  -23.934 &  -23.537 \\
  0.03 &    3.815 &  -20.880 &  -20.698 &  -20.471 \\
  0.04 &    3.003 &  -18.930 &  -18.807 &  -18.652 \\
  0.05 &    2.440 &  -17.634 &  -17.541 &  -17.423 \\
  0.06 &    2.023 &  -16.698 &  -16.623 &  -16.527 \\
  0.07 &    1.701 &  -15.985 &  -15.922 &  -15.841 \\
  0.08 &    1.443 &  -15.421 &  -15.366 &  -15.295 \\
  0.09 &    1.233 &  -14.964 &  -14.915 &  -14.851 \\
  0.10 &    1.058 &  -14.584 &  -14.539 &  -14.481 \\
 \hline
 \end{tabular}
 \end{table}

In the limit of zero energy of the emitted gluon, both invariants
$y_{13}$ and $y_{23}$ approach zero and the first term
in the \eq{eq:t1} develops an IR singularity.
The IR-finite three-jet decay width of the $Z$-boson at LO is given by
the integral of the transition probability $T_{b\bar{b}g}(y_{13},y_{23})$,
\eq{eq:t1}, over the {\it three-jet} region of the phase-space, 
where the soft-gluon region 
is excluded:
\begin{equation}
\label{eq:gammabbgh}
\Gamma^b_{3j} = \left[\frac{1}{2\sqrt{s}}
\int dPS(3) \overline{\sum} \mid M_{b\bar{b}g} \mid^2
 \theta_c(\yc;y_{13},y_{23})
\right]_{\epsilon =0,~s=m_Z^2}~.
\end{equation}
The function $\theta_c(\yc;y_{13},y_{23})$ 
introduces the appropriate cuts for each of the jet-clustering algorithms
and defines the three-jet region.

At the lowest order, E, JADE and EM give
the same three-jet rates for massless particles, because in this case
there is no parton recombination involved and all the three schemes have
the same resolution parameter. However, already at order $\as^2$ they give 
different results since after the first recombination the 
pseudo-particles are not massless anymore and the resolution functions
are different. 
For massive quarks the three algorithms, E, JADE and EM are
already different at order $\as$. The DURHAM algorithm
is, of course, completely different from the other
algorithms we use, both in the massive and the massless cases.

Tables \ref{tab:e0}, \ref{tab:em0}, \ref{tab:jade0} and \ref{tab:durham0}
present the dependence of the function $A^{(0)}(\yc)$ and the ratio
of functions $B\el{0}_{V(A)}(\yc,\rb)/A^{(0)}(\yc)$ on the jet-resolution
parameter $\yc$ for the different jet schemes and for several 
relevant values of $M_b$. It is clearly seen from the tables that 
both leading order functions $B\el{0}_{V}$ 
and $B\el{0}_{A}$ have a soft dependence on $M_b$
for the region of y-cut, $\yc \ge 0.01$, which is relevant for experimental 
studies.

\section{Three-jet ratios at next-to-leading order: overview of the
calculation}
\label{sec:nlo}

Like for many QCD processes it turns out that 
the leading order calculation of the $Z$-boson decay into 
three-jets, described in the previous section, 
does not match the experimental precision.
The numerical results for the leading order calculation were presented 
in~\cite{Ballestrero:1992ed,Ballestrero:1994dv,Bilenkii:1995ad} (see also 
section~\ref{sec:3jets0}).
It is known~\cite{Bethke:1992wk} that next-to-leading corrections 
to the decay of the $Z$-boson  into massless quarks
are very significant numerically.
In addition, at the leading order we do not have any idea what value of the 
quark mass should be taken.
The various perturbative definitions of the quark mass 
(pole mass, running mass etc.) differ only
at the next-to-leading order. However, the numerical difference  
when several mass definitions are used in the leading order 
prediction for three-jet observables is large~\cite{Bilenkii:1995ad}. 
It is again an indication
that next-to-leading corrections are important.
Therefore, to use consistently the quark mass definition,
and to provide accurate predictions for three-jet observables, 
the next-to-leading order, ${\cal O}(\alpha_s^2)$, calculation
for massive quarks has to be included. 

The main difficulty of the NLO calculation is 
the appearance at the intermediate stages, 
in addition to ultraviolet divergences,
of infrared and collinear singularities due to massless gluons.
The Bloch-Nordsieck and Kinoshita-Lee-Nauenberg
theorems~\cite{Bloch:1937pw,Kinoshita:1962ur,Lee:1964is}
assure, however, that observable jet cross-sections 
are infrared finite and free from collinear divergences. 

At the NLO the three-jet cross-section has contributions 
from three-parton, figs.~\ref{fig:loops} and \ref{fig:self}, and four-parton,
figs.~\ref{fig:bbgg}, \ref{fig:bbqq} and \ref{fig:bbbb}, final states.

The IR and collinear singularities of the NLO one-loop Feynman diagrams
cancel against divergences that appear
when the differential cross-section for four-parton production
is integrated over the region of phase-space where either one gluon is soft 
or two gluons are collinear. 
However, the singularities that appear in the intermediate steps 
of the calculation should be treated properly. 
We used dimensional regularization to
regularize both UV and IR 
divergences~\cite{tHooft:1972fi,Marciano:1975tv,Gastmans:1976sr}
because this regularization preserves the QCD Ward identities.

The three-parton transition amplitude can be expressed in terms of a few
scalar one-loop integrals. The result contains poles in 
$\epsilon=(4-D)/2$, where $D$ is a space-time dimension.
Some of the poles have UV origin and other correspond to the IR singularities.
All UV divergences are removed by appropriate renormalization 
of the parameters  of the QCD Lagrangian (coupling constant, 
mass and wave functions).
After UV renormalization and taking the interference with the tree-level 
three-parton amplitude
we obtain analytical expressions which have a part containing the IR poles 
(single and double) and a finite contribution. The singular part
is proportional to the leading order transition 
probability.

The four-parton transition probabilities are split in two parts.
The first one, the so-called soft part, contains singularities
when one of the massless partons in the final state
is soft, or two massless final partons are collinear. The second part,
denoted as hard, is free from any potential singularities.
Then, to cancel the IR/collinear divergences against
those appearing in the three-parton
contribution we work in the context of the
phase-space slicing method~\cite{\cslice}.
In this method the analytical integration over a thin slice at the
border of the phase-space, which contains the IR/collinear singularities,
is performed in $D$-dimensions by using approximate expressions for the 
transition probability expanded in the slice-parameter that defines the 
size of the slice.
The result is added to the virtual corrections. The sum becomes free
of singularities and can be integrated numerically for $D=4$ in the 
three-jet region of the three-body phase-space 
(defined by the jet-clustering algorithms considered in the previous section: 
EM, JADE, E and DURHAM),
but it depends on the
small slice-parameter. The integration of the hard part
over the three-jet region of the four-body phase-space
(again defined by the jet-clustering algorithms considered above)
is done numerically for $D=4$. 

The sum of
the three-parton and four-parton contributions to the
three-jet decay width of the $Z$ boson
should be independent of the parameter that defines 
the slice as long as it is small enough not to spoil the approximations
used. The independence of this parameter has been checked in our calculations. 
Finally, we obtain the finite functions
$H_V(\yc,\rb)$ and $H_A(\yc,\rb)$ in \eq{eq:Bform} at order
$\as$. 

Our classification of the transition probabilities is similar to 
the one used by Ellis, Ross and Terrano~\cite{Ellis:1981wv}
which is based on a colour factor classification.
The way the cancellation of IR
divergences occurs can be seen by representing the different
transition probabilities as the different cuts one can perform in the
three-loop bubble diagrams contributing to the $Z$-boson selfenergy.
Therefore, we assign both three- and four-parton
transition probabilities to the different classes defined by the 
six bubble diagrams depicted in fig~\ref{fig:bubbles}.
The advantage of this classification is
that the cancellation of IR divergences occurs for each class of diagrams
separately and, therefore, numerical tests can be performed independently 
for the IR finite result obtained in each class.

\section{Three-parton contributions}
\label{sec:3parton}

\subsection{Classification of diagrams}

The complete set of diagrams describing the one-loop radiative corrections
to the process
\beq
  Z(q) \rightarrow b(p_1) + \bar{b}(p_2) + g(p_3)~,
\eeq
is shown in figs.~\ref{fig:loops} and \ref{fig:self}.
They contribute to the three-jet decay rate
at $O(\as^2)$
through their interference with the lowest order bremsstrahlung diagrams
$Ta$ and $Tb$ in fig.~\ref{fig:zbbg}.
\mafigura{12.cm}{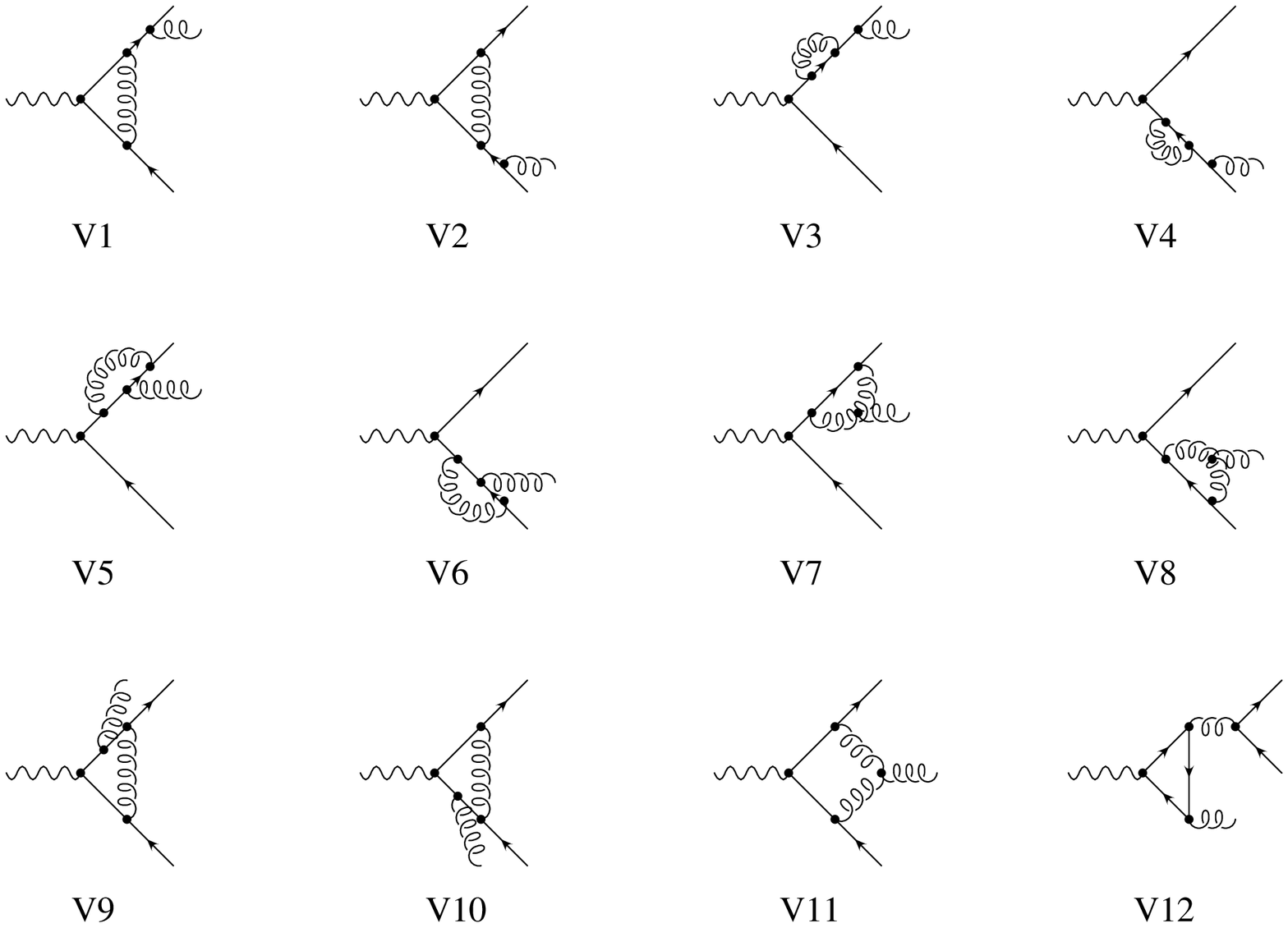}
{Radiative corrections to the process $Z\rightarrow b\bar{b}g$.
Diagrams $V1$ to $V12$ contribute at $O(\as^2)$ through 
their interference with the lowest order bremsstrahlung
diagrams $Ta$ and $Tb$.}
{fig:loops}
\mafigura{12.cm}{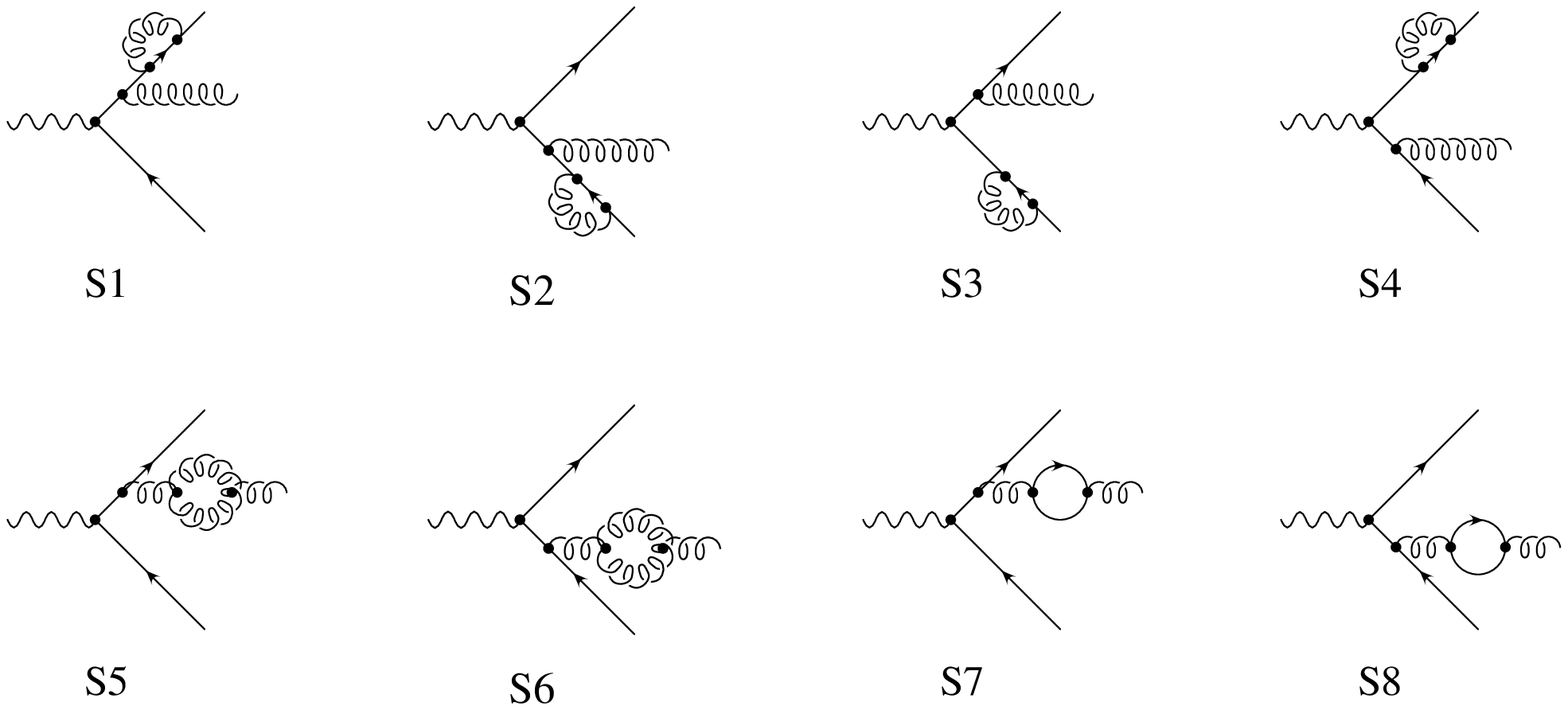}
{Selfenergy diagrams. Graphs involving ghosts 
and similar in structure to $S5$ and
$S6$ have not been shown.}
{fig:self}

We denote by $Via$ and $Sja$ the interference of diagrams $Vi$ and
$Sj$ with $Ta$ averaged (summed) over initial (final) 
colours and spins,
\beq
Via = \overline{\sum} Re \{Vi*Ta^+\}, \qquad
Sja = \overline{\sum} Re \{Sj*Ta^+\}~.
\eeq
Interference with $Tb$ is denoted in the same way by $Vib$ and $Sjb$.

As usual in QCD calculations in the \msb scheme at LEP energies, we are 
working in a theory with $N_F=5$ quark flavours in which only the $b$-quark 
is considered massive.
Therefore, in all closed quark loops
in fig.~\ref{fig:self} all five quarks are running in the loop.
The contribution of the triangle diagram V12 in fig.~\ref{fig:loops} is
peculiar. This contribution is due entirely to the fact that the 
top-quark and the bottom-quark are not degenerated~\cite{Hagiwara:1991dx}
(lighter quarks are effectively degenerated), thus, both top and bottom
quarks are kept in the loop.

\mafigura{13.cm}{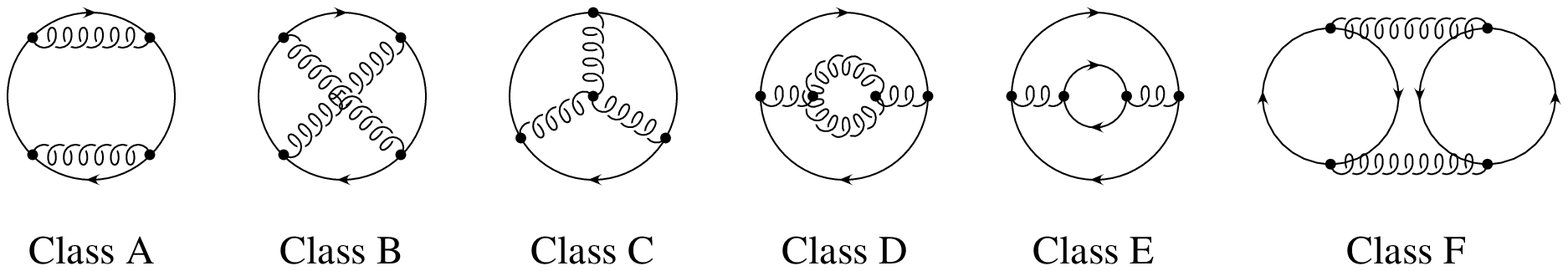}
{Bubble classification relating virtual and real
contributions at $O(\as^2)$ according to their
divergent structure. 
Again we do not show diagrams similar to Class D with 
ghosts in the loop.}
{fig:bubbles}

It turned out very convenient to classify the different one-loop three-parton
as well as four-parton contributions
based on the six three-loop bubble diagrams
presented in fig.~\ref{fig:bubbles}. Different cuts of the same diagram 
represent either a tree-level four-parton transition probability, 
or a one-loop three-parton transition probability, or a two-loop two-parton 
transition probability.
Contributions, corresponding to different bubble diagrams are proportional
to different colour factors. Then, the cancellation of IR 
divergences occurs separately between contributions obtained 
from the same class of diagrams.

The classification of all one-loop contributions to the decay
$Z\rightarrow \bar{b}bg$ is
shown in table~\ref{tab:permutaloop}. In every column of the table we group
contributions with the same colour factor which is shown above the
column. Here $T_R=1/2$.
The transition probabilities in the third and the fourth rows
of the table can be obtained from the ones in the
first and second rows by the interchange of the quark and antiquark momenta.

\begin{table}[hbtp]
\begin{center}
\caption{The interchange table relating the graphs for 
$Z \rightarrow b \bar{b} g$. The third and the fourth rows 
are obtained from the first and the second rows by 
the interchange of the quark and the antiquark momenta
($p_1 \leftrightarrow p_2$). 
All colour factors are given up to a global constant $C_F \: N_C$.
\label{tab:permutaloop}}
\begin{tabular}{|c|c|c|c|c|c|} \hline
Class A & Class B & Class C & Class D & Class E & Class F \\
$C_F$ & $C_F-\frac{1}{2}N_C$ 
& $N_C$ & $N_C$ & $T_R$ & $T_R$ \\
\hline
V1a V3a S1a S3a & V5a V9a  & V7a V11a & S5a & S7a & V12a \\
V1b V3b S1b S3b & V5b V9b  & V7b V11b & S5b & S7b & V12b \\ \hline
V2a V4a S2a S4a & V6a V10a & V8a V11a & S6a & S8a & V12a \\
V2b V4b S2b S4b & V6b V10b & V8b V11b & S6b & S8b & V12b \\
\hline
\end{tabular}
\end{center}
\end{table}

One-loop contributions from Class A and Class B have at most simple IR poles 
for the case of massive quarks. 
This is because the corresponding four-parton contribution
to the three-jet final state can have a singularity only due to the radiation of a 
single soft gluon.

Class C one-loop contributions involving the three-gluon vertex has, in
addition, a double infrared pole. In this case the two gluons in
the corresponding  diagram for the process $Z \rightarrow b\bar{b}gg$
involving the three-gluon vertex can be collinear at the same time
that one of them is soft.

Contributions from Class D 
and Class E with a light quark running
in the inner loop of the bubble
have only simple IR poles coming from the
gluon selfenergy.

Class E with massive quarks in both loops gives a finite result.

Class F results in only a small IR-finite contribution to the axial function
$H_A^{(1)}$ (see \eq{eq:hvaexp}).

The calculational procedure of the one-loop contribution is as follows.
First, we perform all Dirac trace calculations for the matrix element
squared.
Then, as usual, the result is expressed as a sum of Lorentz scalars containing
several loop-integrals with different number of propagators and with
scalar products of the loop momentum and external momenta in the numerator.
We use the standard Passarino-Veltman reduction
procedure~\cite{Passarino:1979jh} to reduce 
all of these vector and tensor loop integrals 
to simpler scalar n-point functions.
Then the final result can be written analytically in terms of
two IR divergent four-propagator (box) integrals, five
three-propagator integrals and a number of simple two-point and one-point 
scalar integrals.
The analytical results for these scalar loop integrals are summarized in
appendix~\ref{ap:loop}.

As mentioned above, after dimensional regularization,  both UV and IR 
singularities appear as poles $\epsilon$. The next step is renormalization
and cancellation of UV divergences.

\subsection{Renormalization counterterms}

Renormalization is accomplished by using a mixed scheme, on-shell
for quark masses and \msb for the gauge coupling. Then, to subtract
the UV divergences we make the following replacements
in the Born decay widths,
\bea
M_b & \rightarrow & M_b
\left\{ 1 - C_F \frac{\as}{4\pi} 
\left[ 3 \left( 
\Delta - \log \frac{M_b^2}{\mu^2} \right) + 4 
\right]  \right\}~, \\
g_s & \rightarrow & g_s \left\{ 1+ \frac{\as}{4\pi} 
\left[ \frac{11}{6} N_C - \frac{3}{2} T_R N_F \right] 
\Delta \right\}~,
\eea
where $\Delta=(4\pi)^{\epsilon}/(\epsilon \: \Gamma(1-\epsilon))$.
Thus, the counterterm that should be added to the one-loop decay width
to eliminate all the UV divergences has the form,
\beq
\Gamma^{CT} = 2 \: \bar{\Gamma}^0 \:
\as^2 \: \left[ \frac{3}{2} T_R N_F
- \frac{11}{6} N_C \right] \frac{1}{\epsilon} 
\int dPS(3) \: T_{b\bar{b}g}(y_{13},y_{23})~,
\eeq
where we defined,
\beq
\label{eq:gammadefs}
\Gamma^0 = \frac{1}{2\sqrt{s}} \frac{g^2}{16 c_W^2}
\frac{C_F N_C}{3-2\epsilon}, \qquad 
\bar{\Gamma}^0 = \Gamma^0 
\mu^{2\epsilon}s^{-\epsilon} 
\frac{(4\pi)^{\epsilon}}{\Gamma(1-\epsilon)}~, 
\eeq
and $T_{b\bar{b}g}(y_{13},y_{23})$ is the square of the tree-level single
bremsstrahlung amplitude, see~\eq{eq:t1}.

After renormalization has been performed, 
the three-parton differential transition probability has a finite part
and divergent terms proportional to the IR poles.

\subsection{Infrared divergent contributions}

The IR divergent piece of the 
one-loop contribution to the decay width
$Z \rightarrow \bar{b}bg$ reads
(See appendix \ref{ap:loop} for notation and details):
\bea
\Gamma^{(s)}_{virtual} &=& 2 \bar{\Gamma}^0
\: \as^2  \: 
\int dPS(3) \: T_{b\bar{b}g}(y_{13},y_{23})   
\non \\ &\times&
\left\{ - \frac{2}{\epsilon} C_F  
- \frac{2}{\epsilon} \left(C_F-\frac{N_C}{2}\right) 
\frac{y_{12}}{(y_{12}+2r_b)\beta_{12}} \log c_{12}
\right. \non \\ & & 
- N_C \left[ \frac{1}{\epsilon^2}
+ \frac{1}{\epsilon} \left( \frac{11}{6} 
+ \log \frac{r_b}{y_{13} \: y_{23}} \right) 
\right] 
\non \\ & & \left. 
+ \frac{2}{3} T_R \left[ (N_F-1) \frac{1}{\epsilon} + \log r_b \right]
\right\}~.
\label{eq:divvirtual}
\eea
Note that in \eq{eq:divvirtual} we also included a term proportional
to $T_R$ and $\log{r_b}$, which is finite for a non-zero quark mass
($r_b \ne 0$). This term comes from the
gluon self-energy and would appear as an additional pole if
the $b$-quark had been considered massless. It is canceled by
a four-parton contribution in which a gluon emitted from the initial 
$b$-quarks splits into two heavy quarks. It is important to note that
similar contributions appear in the case of gluons emitted from
light quarks. 

\section{Four-parton contributions}
\label{sec:4parton}

\subsection{Classification of Diagrams}

\subsubsection{Emission of two real gluons}

The process
\beq
Z(q) \rightarrow b(p_1) + \bar{b}(p_2) + g (p_3) + g (p_4)~
\end{equation}
is described by the eight diagrams shown in fig.~\ref{fig:bbgg} and, 
therefore, the transition probability contains, in principle, 36 contributions.
However, many of them are related
by interchange of momentum labels and, at the end, only 13
transition probabilities need to be calculated.

\mafigura{12.cm}{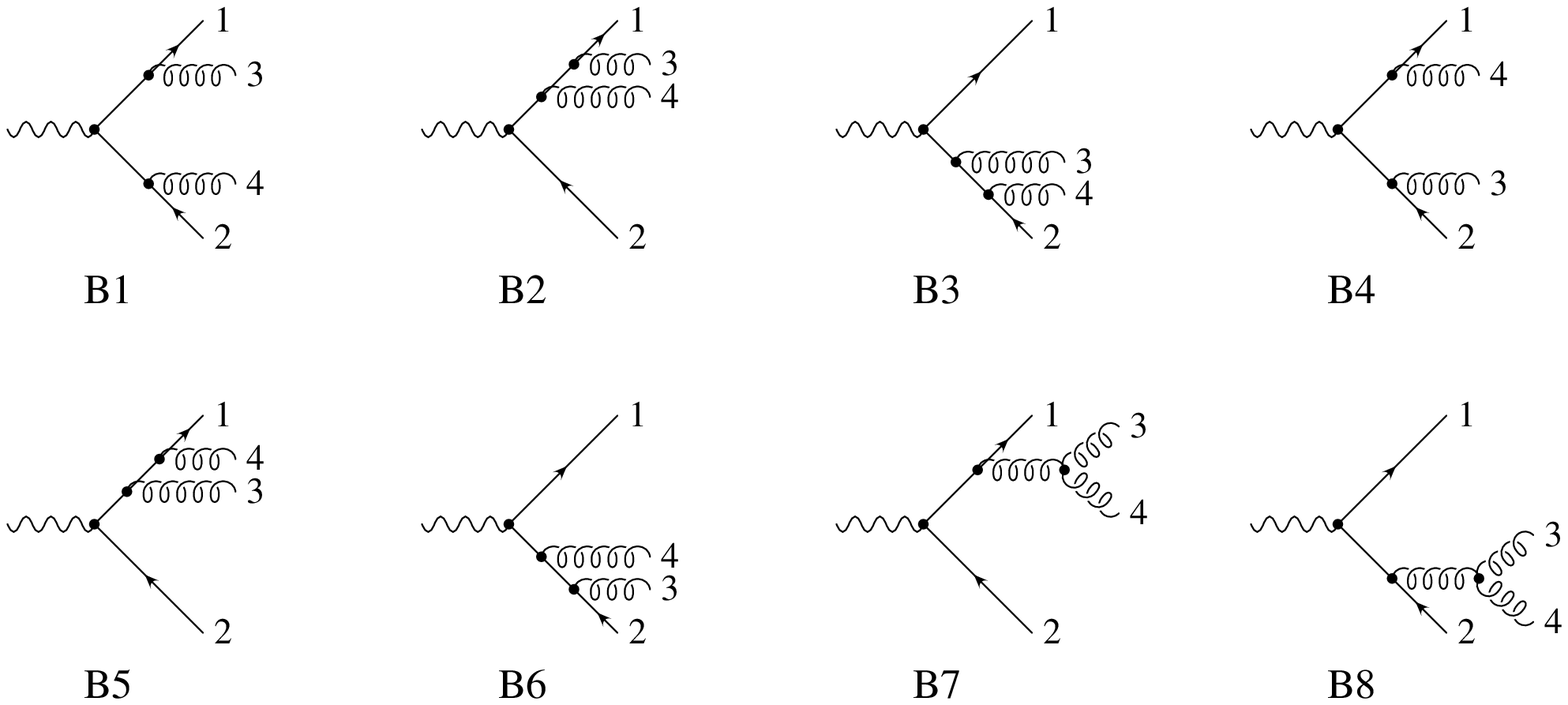}
{Feynman diagrams contributing to the process 
$Z\rightarrow b\bar{b}gg$.}
{fig:bbgg}

As before, we denote by $B_{ij}$
the interference of diagram $B_i$ with $B_j$
averaged (summed) over initial (final) colours and spins, 
\beq
Bij = \overline{\sum} Re \{Bi*Bj^+\}~.
\eeq
In table~\ref{tab:permuta} we give the momentum label 
interchanges necessary to generate all the transition probabilities
and we classify them according to the previously defined bubble groups.
We see from the table that it is sufficient to consider the following
combinations of transition probabilities,
\bea
\mrm{Class A}^{b\bar{b}gg} &=& \frac{1}{2} B11 + 2 B21 + B22 + B32~, \non \\
\mrm{Class B}^{b\bar{b}gg} &=& \frac{1}{2} B41 + 2 B42 + B62 + B52~, \non \\
\mrm{Class C}^{b\bar{b}gg} &=& 2 (B71 + B72+ B82)~, \non \\
\mrm{Class D}^{b\bar{b}gg} &=& \frac{1}{2} (B77+B87)~,
\eea
plus the interchanges $(1 \leftrightarrow 2)$, $(3 \leftrightarrow 4)$
and $(1 \leftrightarrow 2)$ $(3 \leftrightarrow 4)$.

\begin{table}[hbtp]
\begin{center}
\caption{The interchange table relating the graphs for 
$Z \rightarrow b \bar{b} g g$.  
All colour factors are given up to global constant $C_F \: N_C$.
\label{tab:permuta}}
\begin{tabular}{|c|c|c|c|c|} \hline
label       & Class A & Class B & Class C & Class D \\
permutation & $C_F$ & $C_F-\frac{1}{2}N_C$ & $N_C$ & $N_C$ \\
\hline
      & B11 B21 B22 B32 & B41 B42 B62 B52 & B71 B72 B82 & B77 B87 \\
$(1 \leftrightarrow 2)$ 
      & B44 B64 B66 B65 & B41 B61 B62 B63 & B84 B86 B76 & B88 B87 \\
$(3 \leftrightarrow 4)$ 
      & B44 B54 B55 B65 & B41 B51 B53 B52 & B74 B75 B85 & B77 B87 \\
$(1 \leftrightarrow 2)$ $(3 \leftrightarrow 4)$ 
      & B11 B31 B33 B32 & B41 B43 B53 B63 & B81 B83 B73 & B88 B87 \\
\hline
\end{tabular}
\end{center}
\end{table}

The sum over the two physical polarizations of the
produced gluons is accomplished by summing over
the polarizations with,
\beq
\sum_{pol} \varepsilon^{\mu *} \varepsilon^{\nu} 
= - g^{\mu \nu}~,
\end{equation}
but including in Class~D Feynman diagrams like $B7$ and $B8$ 
with ``external'' ghosts in order to take into account the
fact that the gluon current is not conserved.   

Only gluons attached to external legs (quarks or gluons) can generate
infrared divergences in the three-jet region. 
Thus, $B32$ and $B52$ are fully finite, $B21$, $B22$, $B42$ and $B62$ are
IR divergent only if gluon labeled as 3 is soft
while $B11$ and $B41$ are IR divergent in the three-jet region when
any of the gluons labeled as 3 or 4 is soft. 
As commented before, this kind of transition probabilities, classified 
into Classes A and B, contain only simple infrared poles 
in $\epsilon$ since for massive quarks the quark-gluon collinear
divergences are softened into logarithms of the quark mass.
On the other hand, double infrared poles appear 
for diagrams of Class C because the gluon-gluon collinear
divergences are present due to the three-gluon vertex.
The situation is similar for $B77$ and $B87$. These diagrams
individually contain double infrared poles. Nevertheless,
their sum is such that at the end only simple infrared poles survive. 
The reason being that they belong 
to the same Class~D as the diagrams with a selfenergy insertion
in an external gluon leg.

\subsubsection{Emission of four quarks}

First we consider the process
\bea
Z(q) & \rightarrow b(p_1) + \bar{b}(p_2) + q(p_3) + \bar{q}(p_4)~,
\eea
where $q$ stands for a light quark, which is assumed massless.

\mafigura{6cm}{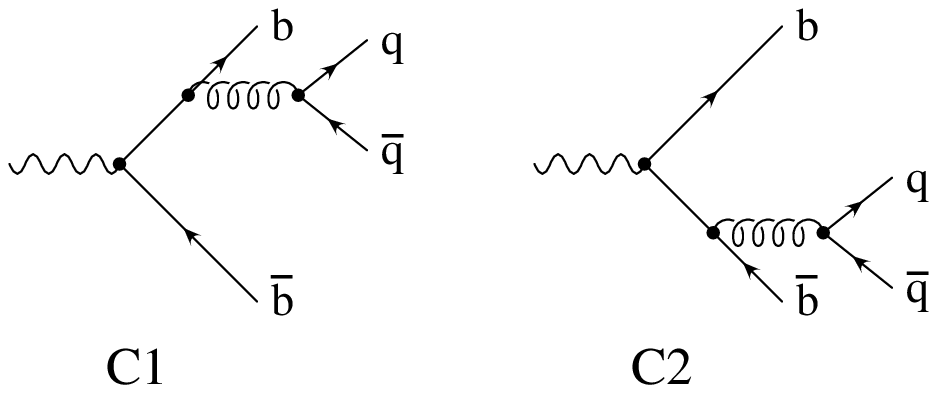}
{Feynman diagrams contributing to the process
$Z \rightarrow b \bar{b} q \bar{q}$, where $q$ stands for a
light quark.}
{fig:bbqq}

This process is described by diagrams shown in the fig.~\ref{fig:bbqq},
where the pair of bottom-antibottom quarks is emitted from the primary
vertex, and two similar Feynman diagrams, where the heavy quark pair is
radiated off a light $q\bar{q}$ system attached to the $Z$-boson (the
so-called heavy quark gluon splitting diagrams).

The transition probabilities $C11$, $C22$ and $C12$, corresponding to
the diagrams in fig.~\ref{fig:bbqq}, are assigned to Class~E
\beq
\mrm{Class E}^{b\bar{b}q\bar{q}} = C11 + C12 + (1 \leftrightarrow 2)~,
\eeq
and can be obtained from the bubble diagram, fig.~\ref{fig:bubbles},
with a heavy quark in the external loop and a light quark in the internal loop.
Due to the light quarks, which are considered massless,
the above transition probabilities have
soft and quark-quark collinear singularities in the three-jet region.
In principle, these divergences can contain
double poles\footnote{For exactly massless quarks. It is also
possible to regularize these infrared divergences
with a small quark mass. In this case, infrared divergences
are softened into mass singularities and lead to
large logarithms in the quark mass, $\log(m_q/\mu)$.
Similarly, infrared gluon divergences can be regulated at lowest order
by giving a small mass, $\lambda$, to the gluons.
At next-to-leading order we would violate gauge invariance
at the three-gluon vertex.}.
However, as in the case of the soft and the gluon-gluon collinear divergences
in $B77$, $B88$ and $B87$, all double
poles cancel in the sum, because these transition probabilities
are related to the one-loop three-parton contributions with
a external gluon self-energy insertion (they correspond to different cuts of
the same bubble diagram), which has only a simple IR pole.

The remaining contributions to the decay $Z \rightarrow
\overline{b}b\overline{q}q$ correspond to the heavy quark gluon splitting
diagrams and to their interference with the diagrams shown in 
fig.~\ref{fig:bbqq}. They can be understood
following the bubble representation discussed in the previous section and
depicted in fig.~\ref{fig:bubbles}.

The contribution of the heavy quark gluon splitting diagrams is IR finite.
It is obtained from Class E bubble diagrams in fig.~\ref{fig:bubbles} with a
light quark (external loop) and a heavy quark (internal loop). This
four-parton contribution contains numerically large collinear logarithms of
the heavy quark mass. However, they all cancel if the four-parton part is
summed together with the one-loop three-parton contributions with light
quarks coupled to the $Z$ boson and a gluon self-energy insertion with a
heavy quark in the loop. These three-parton contributions are obtained from
the same Class E bubble diagram by cutting the gluon and the light quark
lines. In principle, one can assign the heavy quark quark gluon splitting
contributions to either the light quark three-jet decay width or to the
heavy quark three-jet decay width. Because the cancellations discussed above
we choose to include them in the three-jet $Z$-decay width into light
quarks. In this way the limit $M_b\rightarrow 0$ can be taken and results
can be compared with known massless
calculations~\cite{Kunszt:1989km,Bethke:1992wk,Ellis:1981wv}. Obviously,
these heavy quark gluon splitting contributions should be subtracted from
the experimental data before comparing with our theoretical result.

Finally, we have to consider the contributions corresponding to Class F bubble
diagrams in fig.~\ref{fig:bubbles} which are also IR finite.
By cutting one gluon line and one quark loop in this bubble diagram one obtains
one-loop three-parton contributions arising from the triangle diagrams
$V12$, which, as commented in the previous section, produce a tiny
difference, even for massless quarks, between $H_A(y_c,r_b)$ and
$H_V(y_c,r_b)$. Cutting the two quark loops in Class F bubble diagrams with
one heavy and one light quark loops gives the four-parton contribution from
the interference of diagrams in which a heavy quark pair is radiated off a
light quark pair with diagrams in which a light quark is radiated off a
heavy quark pair. This contribution has the same type of cancellation that
occur for the corresponding three-parton contribution. Numerically it is
even smaller than the one produced by triangle diagrams. Therefore we have
neglected it.

A different approach to avoid the large logarithms has been
followed in \cite{Bernreuther:1997jn,Brandenburg:1997pu},
where a double $b$-tagging is imposed, sacrificing, however,
the experimental statistics.

\mafigura{12.cm}{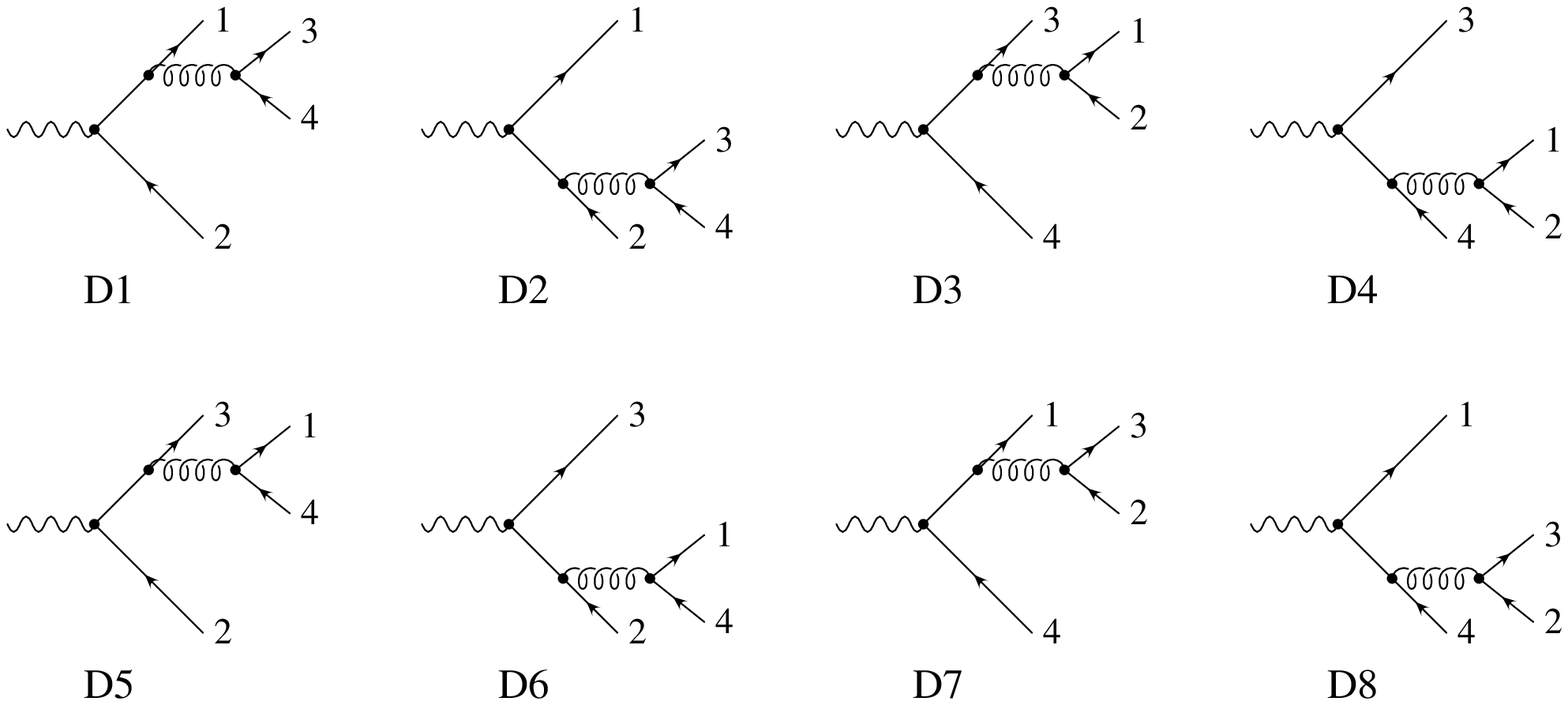}
{Feynman diagrams contributing to the decay width of the
$Z$-boson into four massive bottom quarks.}
{fig:bbbb}

We consider now the decay into four bottom quarks.
\bea
Z(q) & \rightarrow b(p_1) + \bar{b}(p_2) + b(p_3) + \bar{b}(p_4)~,
\eea
The corresponding diagrams are shown in fig.~\ref{fig:bbbb}.
As in the case of the emission of two real gluons,
from the eight diagrams shown in fig.~\ref{fig:bbbb}
we should compute only twelve different terms in the transition probability
out of the 36 terms possible as the other terms are related by
interchange of momentum labels.

\begin{table}[hbtp]
\begin{center}
\caption{The interchange table relating the graphs for 
$Z \rightarrow b \bar{b} b \bar{b}$.A global $C_F \: N_C$ factor 
has been factorized from the colour factor.\label{tab:permuta2}} 
\begin{tabular}{|c|c|c|c|} \hline
label       & Class B   & Class E & Class F \\
permutation & $C_F-\frac{1}{2}N_C$ & $T_R$ & $T_R$ \\
\hline
      & D18 D25 D15 D28 D17 D26 & D11 D12 D22 & D13 D23 D24 \\
$(1 \leftrightarrow 3)$ 
      & D45 D16 D15 D46 D35 D62 & D55 D56 D66 & D57 D67 D68\\
$(2 \leftrightarrow 4)$ 
      & D27 D38 D37 D28 D17 D48 & D77 D78 D88 & D57 D58 D68\\
$(1 \leftrightarrow 3)$ $(2 \leftrightarrow 4)$ 
      & D36 D47 D37 D46 D35 D48 & D33 D34 D44 & D13 D14 D24\\
\hline
\end{tabular}
\end{center}
\end{table}

Thus, it is sufficient to consider the following combinations:
\bea
\mrm{Class B}^{b\bar{b}b\bar{b}} 
&=& - (D15 + D28 + D17 + D26 + 2(D18 + D25))~, \non \\
\mrm{Class E}^{b\bar{b}b\bar{b}} &=& D11 + 2 D12 + D22~, \non \\
\mrm{Class F}^{b\bar{b}b\bar{b}} &=& D13 + 2 D23 + D24~,
\eea
plus the interchanges $(1 \leftrightarrow 3)$, $(2 \leftrightarrow 4)$
and $(1 \leftrightarrow 3)$ $(2 \leftrightarrow 4)$.
Due to Fermi statistics there is a relative minus sign for diagrams
$D5$ to $D8$ that is reflected in the transition probabilities
of Class B$^{b\bar{b}b\bar{b}}$ and ensures they vanish 
when both fermions (antifermions) have identical quantum numbers.

The transition probabilities of Class~F
are called in the literature~\cite{Chetyrkin:1994js}
singlet contributions
because they contain two different fermion loops and hence 
can be split into two parts by cutting gluon lines only.
The first contribution to the vectorial part arises 
at $O(\as^3)$ as a consequence of the non-abelian 
generalization of Furry's theorem.
Singlet contributions to the axial part appear already 
at $O(\as^2)$. 

\subsection{Infrared divergent contributions}

The part of the four-parton decay width which is
singular when a gluon (with momentum $p_3$) attached to a massive quark 
becomes soft is given by,
\bea
\Gamma_{I}^{(s)} &=& 
2 \Gamma^0
\: g_s^4 \: \frac{1}{2!s} \int dPS(3)(y_{14},y_{24})
T_{b\bar{b}g}(y_{14},y_{24})
\non \\ &\times &
\frac{\mu^{2\epsilon}}{2(2\pi)^{D-1}}
\int_0^{w\sqrt{s}} E_3^{D-3} dE_3 d\Omega_3
\left\{ \left(C_F-\frac{N_C}{2}\right) \frac{y_{12}}{y_{13}\: y_{23}}
- C_F \frac{2 r_b}{y_{13}^2} + O(E_3^{-1}) \right\}
\non \\
&+& (1\leftrightarrow 2) + (3\leftrightarrow 4)
+ (1\leftrightarrow 2)(3\leftrightarrow 4)~.
\eea

The part of the four-parton decay, corresponding to the diagrams with
a gluon splitting into two gluons or into two light quarks, which is
singular when one of the two gluons (or light quarks) is
soft, or two massless partons are collinear, is given by
(See appendices \ref{ap:ps} and \ref{ap:divs} for notation and details),
\bea
&&\Gamma_{II}^{(s)} = 2 \bar{\Gamma}^0 \:
\as^2  \: \int dPS(3)(y_{134},y_{234})
\non \\ 
&&\times 
\int_0^w y_{34}^{-1-\epsilon} dy_{34} \int_0^1 dv (v(1-v))^{-\epsilon}
\frac{1}{N_{\theta'}} \int_0^{\pi} d\theta' \sin^{-2\epsilon} \theta'
\non \\ 
&& \times  
\left[T_{b\bar{b}g}(y_{134},y_{234}) 
\left\{ \frac{N_C}{2!} \left( 
\frac{y_{134}}{y_{13}} + \frac{y_{234}}{y_{23}}
+ \frac{y_{134}}{y_{14}} + \frac{y_{234}}{y_{24}}  
- 4 + 2v(1-v) \right)
\right. \right. \non \\ 
&&\left. \qquad \qquad \qquad \qquad 
+ T_R(N_F-1) \frac{v^2+(1-v)^2-\epsilon}{1-\epsilon} \right\} 
\non \\ 
&& \quad
+ 32 \frac{h_p}{y_{13}^2 y_{23}^2} T_{b\bar{b}} v(1-v) 
\left( 2\cos^2 \theta'-\frac{1}{1-\epsilon}\right) \non \\
&& \left. \qquad \qquad \qquad \qquad \times 
\left\{ \frac{N_C}{2!} (1-\epsilon) - T_R(N_F-1) \right\}
+ O(y_{34}^0) \right]~.
\eea
After integration over the angle $\theta'$ the last term,
proportional to the lowest order transition probability $T_{b\bar{b}}$,
vanishes. The other terms become, as expected, proportional 
to the well-known Altarelli-Parisi kernels for $v \neq 0,1$
\bea
P_{gg}(v) &=& N_C \left[ \frac{2}{v} + \frac{2}{1-v}
- 4 + 2v(1-v)  \right] \non \\
P_{qg}(v) &=& T_R \left[ \frac{v^2+(1-v)^2-\epsilon}{1-\epsilon} \right]~.
\eea
This result follows straightforwardly from the fact that in the limit
considered we have
$y_{13}\rightarrow y_{134} v$ and similar behaviour for  
other kinematical invariants. 
Using the phase-space integrals presented in 
appendix~\ref{ap:divs} we obtain the IR divergent 
contribution in the three-jet region of the four-parton 
processes,
\bea
\Gamma^{(s)}_{real} &=& 2 \bar{\Gamma}^0
\: \as^2  \: 
\int dPS(3) \: T_{b\bar{b}g}(y_{13},y_{23})   
\non \\ &\times&
\left\{\frac{2}{\epsilon} C_F  
+ \frac{2}{\epsilon} \left(C_F-\frac{N_C}{2}\right) 
\frac{y_{12}}{(y_{12}+2r_b)\beta_{12}} \log c_{12}
\right. \non \\ & & 
+ N_C \left[ \frac{1}{\epsilon^2}
+ \frac{1}{\epsilon} \left( \frac{11}{6} 
+ \log \frac{r_b}{y_{13} \: y_{23}} \right) 
\right] 
\non \\ & & \left. 
- \frac{2}{3} T_R \left[ (N_F-1) \frac{1}{\epsilon} + \log r_b \right]
\right\}~,
\label{eq:divreal}
\eea
where, for consistency, we have also included a $\log \rb$ term 
coming from the integration of four massive quark transition 
probabilities. 
From \eq{eq:divvirtual}, \eq{eq:divreal} we can
easily see that, as expected, the IR divergences cancel between 
three-parton and four-parton contributions rendering the final answer 
completely finite. In fact, one can see that
this cancellation occurs separately for the different groups of diagrams
defined in fig.~\ref{fig:bubbles}.

\section{Results and applications: $R_3^{b\ql}$, $D_2^{b\ql}$ and 
$m_b(m_Z)$ from data at the $Z$ peak}
\label{sec:mb}

Since a big part of the calculation has been done numerically, it is important
to have some checks of it. We have checked our 
transition probabilities for four-parton final states in 
the massless limit
against the ones calculated by 
Ellis, Ross and Terrano (ERT)~\cite{Ellis:1981wv}.
The massless limit cannot be taken directly 
in the three-parton transition probabilities, since, 
as commented before, some collinear poles in $\epsilon$ appear in our 
calculation as logarithms of the heavy quark mass.
As seen before, we have checked that all the IR divergences cancel between
three-parton and four-parton contributions in the massive case.
To check the performance of the numerical procedure we integrated the
massless amplitudes of ERT and obtained the known results for the
functions $A\el{1}$. In addition 
our four-parton amplitudes have been checked in the case of massive
quarks, in four dimensions, by comparing their contribution to four-jet
processes to the known results~\cite{Ballestrero:1992ed,Stelzer:1994ta}.
Finally, we have checked, independently for each class of diagrams
with different colour factors, that the final result obtained with massive 
quarks reduces to the massless result in the limit of very small masses.

The last check is the main check of our calculation.
We have checked numerically that in the limit of $M_b \rightarrow 0$
we recover the values already known for the functions $A\el{0}(\yc)$ and
$A\el{1}(\yc)$ in  the different
algorithms considered. We did so by computing the functions $H_V(\yc,\rb)$ and
$H_A(\yc,\rb)$ for several small values of $\rb$ and then extrapolating
the results for $\rb \rightarrow 0$. This check is not trivial at all
due to the already emphasized difference in the IR structure of the 
massless and massive cases~\cite{Rodrigo:1996gw}.

Finally we have compared numerically some partial results with the
results obtained in ~\cite{\cnason}, where 
a completely different approach has been used to cancel IR divergences, 
and have found good agreement.

 \begin{table}[htb]
  \caption{The next-to-leading order functions 
  (see eqs.~(\ref{eq:gamma3jets}, \ref{eq:hvaexp}, \ref{eq:Bform})) for
  the E algorithm.\label{tab:e}}
 \begin{tabular}{|c|r|r|r|r|}
 \hline
 $y_c$   & $A^{(1)}_V \: \:$ &
 \multicolumn{3}{c|}{$B^{(1)}_V/A^{(1)}_V$}
 \\\cline{3-5}&
 & $M_b=3$GeV & $M_b=4$GeV & $M_b=5$GeV
 \\ \hline
$0.01$  & $   165.014(156)$ & $ 874.1 \pm   1.4 $    
        & $ 467.5 \pm   0.7 $ & $ 245.7 \pm   0.4 $  \\
$0.02$  & $   153.145( 85)$ & $ 411.8 \pm   1.0 $    
        & $ 288.9 \pm   0.5 $ & $ 193.8 \pm   0.3 $  \\
$0.03$  & $   128.583( 64)$ & $ 221.4 \pm   1.0 $    
        & $ 192.4 \pm   0.5 $ & $ 149.0 \pm   0.3 $  \\
$0.04$  & $   107.624( 53)$ & $ 136.1 \pm   1.0 $    
        & $ 130.2 \pm   0.5 $ & $ 114.2 \pm   0.3 $  \\
$0.05$  & $    90.606( 46)$ & $  99.0 \pm   1.1 $    
        & $  92.9 \pm   0.6 $ & $  87.6 \pm   0.3 $  \\
$0.06$  & $    76.839( 40)$ & $  78.2 \pm   1.2 $    
        & $  72.5 \pm   0.6 $ & $  68.6 \pm   0.4 $  \\
$0.07$  & $    65.565( 36)$ & $  65.3 \pm   1.2 $    
        & $  60.7 \pm   0.7 $ & $  56.4 \pm   0.4 $  \\
$0.08$  & $    56.198( 32)$ & $  57.4 \pm   1.4 $    
        & $  52.8 \pm   0.8 $ & $  48.6 \pm   0.4 $  \\
$0.09$  & $    48.344( 30)$ & $  50.4 \pm   1.4 $    
        & $  47.0 \pm   0.7 $ & $  43.0 \pm   0.5 $  \\
$0.10$  & $    41.701( 27)$ & $  45.9 \pm   1.8 $    
        & $  42.1 \pm   0.9 $ & $  38.9 \pm   0.5 $  \\
 \hline
 $y_c$   & $A^{(1)}_A-A^{(1)}_V$ &
 \multicolumn{3}{c|}{$B^{(1)}_A/A^{(1)}_V$}
 \\\cline{3-5}&
 & $M_b=3$GeV & $M_b=4$GeV & $M_b=5$GeV
 \\ \hline
$0.01$  & $   0.97579( 21)$ & $ 868.6 \pm   1.4 $    
        & $ 460.6 \pm   0.7 $ & $ 239.7 \pm   0.4 $  \\
$0.02$  & $   0.89814( 18)$ & $ 406.9 \pm   1.0 $    
        & $ 283.6 \pm   0.5 $ & $ 188.2 \pm   0.3 $  \\
$0.03$  & $   0.82866( 15)$ & $ 217.7 \pm   0.9 $    
        & $ 188.2 \pm   0.5 $ & $ 144.0 \pm   0.3 $  \\
$0.04$  & $   0.76466( 13)$ & $ 133.0 \pm   1.0 $    
        & $ 126.5 \pm   0.5 $ & $ 109.8 \pm   0.3 $  \\
$0.05$  & $   0.70563( 12)$ & $  95.8 \pm   1.1 $    
        & $  89.5 \pm   0.6 $ & $  83.7 \pm   0.3 $  \\
$0.06$  & $   0.65020( 11)$ & $  75.3 \pm   1.2 $    
        & $  69.5 \pm   0.6 $ & $  65.0 \pm   0.4 $  \\
$0.07$  & $   0.59832( 11)$ & $  62.5 \pm   1.2 $    
        & $  57.6 \pm   0.6 $ & $  53.2 \pm   0.4 $  \\
$0.08$  & $   0.54944( 10)$ & $  54.1 \pm   1.4 $    
        & $  49.7 \pm   0.8 $ & $  45.4 \pm   0.4 $  \\
$0.09$  & $   0.50324(  9)$ & $  47.6 \pm   1.4 $    
        & $  44.0 \pm   0.7 $ & $  39.9 \pm   0.5 $  \\
$0.10$  & $   0.45975(  9)$ & $  42.7 \pm   1.8 $    
        & $  39.4 \pm   0.9 $ & $  36.1 \pm   0.5 $  \\
 \hline
 \end{tabular}
 \end{table}

 \begin{table}[htb]
  \caption{The next-to-leading order functions 
  (see eqs.~(\ref{eq:gamma3jets}, \ref{eq:hvaexp}, \ref{eq:Bform})) for
  the EM algorithm.\label{tab:em}}
 \begin{tabular}{|c|r|r|r|r|}
 \hline
 $y_c$   & $A^{(1)}_V \: \:$ &
 \multicolumn{3}{c|}{$B^{(1)}_V/A^{(1)}_V$}
 \\\cline{3-5}&
 & $M_b=3$GeV & $M_b=4$GeV & $M_b=5$GeV
 \\ \hline
$0.01$  & $   139.008(152)$ & $  -4.5 \pm   1.0 $    
        & $   1.6 \pm   0.5 $ & $   4.4 \pm   0.3 $  \\
$0.02$  & $   131.434( 82)$ & $ -21.1 \pm   0.8 $    
        & $ -15.2 \pm   0.4 $ & $ -11.8 \pm   0.2 $  \\
$0.03$  & $   109.542( 62)$ & $ -19.4 \pm   0.8 $    
        & $ -16.0 \pm   0.4 $ & $ -13.3 \pm   0.2 $  \\
$0.04$  & $    90.576( 51)$ & $ -16.1 \pm   0.9 $    
        & $ -14.5 \pm   0.5 $ & $ -13.2 \pm   0.3 $  \\
$0.05$  & $    75.182( 44)$ & $ -13.7 \pm   1.0 $    
        & $ -12.3 \pm   0.5 $ & $ -11.9 \pm   0.3 $  \\
$0.06$  & $    62.773( 38)$ & $ -11.6 \pm   1.1 $    
        & $ -11.0 \pm   0.6 $ & $ -10.7 \pm   0.3 $  \\
$0.07$  & $    52.685( 35)$ & $ -10.4 \pm   1.2 $    
        & $ -10.0 \pm   0.6 $ & $ -10.0 \pm   0.4 $  \\
$0.08$  & $    44.354( 31)$ & $  -9.9 \pm   1.3 $    
        & $  -9.4 \pm   0.7 $ & $  -9.2 \pm   0.4 $  \\
$0.09$  & $    37.443( 28)$ & $ -10.2 \pm   1.5 $    
        & $  -9.0 \pm   0.7 $ & $  -8.9 \pm   0.4 $  \\
$0.10$  & $    31.650( 26)$ & $  -8.8 \pm   1.6 $    
        & $  -8.2 \pm   0.9 $ & $  -8.3 \pm   0.5 $  \\
 \hline
 $y_c$   & $A^{(1)}_A-A^{(1)}_V$ &
 \multicolumn{3}{c|}{$B^{(1)}_A/A^{(1)}_V$}
 \\\cline{3-5}&
 & $M_b=3$GeV & $M_b=4$GeV & $M_b=5$GeV
 \\ \hline
$0.01$  & $   0.97579( 21)$ & $  -4.8 \pm   1.0 $    
        & $   0.7 \pm   0.5 $ & $   2.8 \pm   0.3 $  \\
$0.02$  & $   0.89814( 18)$ & $ -22.4 \pm   0.8 $    
        & $ -16.6 \pm   0.4 $ & $ -13.4 \pm   0.2 $  \\
$0.03$  & $   0.82866( 15)$ & $ -20.8 \pm   0.8 $    
        & $ -17.7 \pm   0.4 $ & $ -15.3 \pm   0.2 $  \\
$0.04$  & $   0.76466( 13)$ & $ -17.9 \pm   0.9 $    
        & $ -16.3 \pm   0.5 $ & $ -15.1 \pm   0.3 $  \\
$0.05$  & $   0.70563( 12)$ & $ -15.6 \pm   1.0 $    
        & $ -14.4 \pm   0.5 $ & $ -13.8 \pm   0.3 $  \\
$0.06$  & $   0.65020( 11)$ & $ -13.6 \pm   1.1 $    
        & $ -12.8 \pm   0.6 $ & $ -12.6 \pm   0.3 $  \\
$0.07$  & $   0.59832( 11)$ & $ -12.6 \pm   1.2 $    
        & $ -11.9 \pm   0.6 $ & $ -11.6 \pm   0.4 $  \\
$0.08$  & $   0.54944( 10)$ & $ -12.1 \pm   1.3 $    
        & $ -11.0 \pm   0.7 $ & $ -11.0 \pm   0.4 $  \\
$0.09$  & $   0.50324(  9)$ & $ -12.8 \pm   1.5 $    
        & $ -10.6 \pm   0.7 $ & $ -10.4 \pm   0.4 $  \\
$0.10$  & $   0.45975(  9)$ & $ -11.8 \pm   1.7 $    
        & $  -9.9 \pm   0.9 $ & $  -9.9 \pm   0.5 $  \\
 \hline
 \end{tabular}
 \end{table}

 \begin{table}[htb]
 \caption{The next-to-leading order functions 
  (see eqs.~(\ref{eq:gamma3jets}, \ref{eq:hvaexp}, \ref{eq:Bform})) for
  the JADE algorithm.\label{tab:jade}}
 \begin{tabular}{|c|r|r|r|r|}
 \hline
 $y_c$   & $A^{(1)}_V \: \:$ &
 \multicolumn{3}{c|}{$B^{(1)}_V/A^{(1)}_V$}
 \\\cline{3-5}&
 & $M_b=3$GeV & $M_b=4$GeV & $M_b=5$GeV
 \\ \hline
$0.01$  & $    22.352(146)$ & $ 638.1 \pm   6.3 $    
        & $ 533.2 \pm   3.2 $ & $ 454.0 \pm   1.8 $  \\
$0.02$  & $    71.356( 78)$ & $  27.7 \pm   1.4 $    
        & $  28.1 \pm   0.7 $ & $  25.4 \pm   0.4 $  \\
$0.03$  & $    72.138( 58)$ & $  -4.7 \pm   1.2 $    
        & $  -1.2 \pm   0.6 $ & $  -0.1 \pm   0.4 $  \\
$0.04$  & $    65.303( 48)$ & $ -12.7 \pm   1.2 $    
        & $  -9.8 \pm   0.6 $ & $  -8.3 \pm   0.4 $  \\
$0.05$  & $    57.366( 41)$ & $ -15.1 \pm   1.3 $    
        & $ -12.6 \pm   0.6 $ & $ -11.4 \pm   0.4 $  \\
$0.06$  & $    49.904( 36)$ & $ -16.7 \pm   1.3 $    
        & $ -13.8 \pm   0.7 $ & $ -12.7 \pm   0.4 $  \\
$0.07$  & $    43.254( 32)$ & $ -16.7 \pm   1.4 $    
        & $ -14.0 \pm   0.7 $ & $ -13.1 \pm   0.4 $  \\
$0.08$  & $    37.438( 28)$ & $ -16.5 \pm   1.6 $    
        & $ -14.7 \pm   0.8 $ & $ -13.5 \pm   0.5 $  \\
$0.09$  & $    32.357( 26)$ & $ -16.3 \pm   1.6 $    
        & $ -13.9 \pm   0.8 $ & $ -13.1 \pm   0.5 $  \\
$0.10$  & $    27.961( 24)$ & $ -16.4 \pm   1.8 $    
        & $ -13.7 \pm   0.9 $ & $ -13.0 \pm   0.6 $  \\
 \hline
 $y_c$   & $A^{(1)}_A-A^{(1)}_V$ &
 \multicolumn{3}{c|}{$B^{(1)}_A/A^{(1)}_V$}
 \\\cline{3-5}&
 & $M_b=3$GeV & $M_b=4$GeV & $M_b=5$GeV
 \\ \hline
$0.01$  & $   0.97579( 21)$ & $ 661.7 \pm   6.3 $    
        & $ 552.2 \pm   3.1 $ & $ 468.1 \pm   1.8 $  \\
$0.02$  & $   0.89814( 18)$ & $  29.8 \pm   1.4 $    
        & $  29.2 \pm   0.7 $ & $  25.7 \pm   0.4 $  \\
$0.03$  & $   0.82866( 15)$ & $  -4.8 \pm   1.2 $    
        & $  -1.4 \pm   0.6 $ & $  -0.6 \pm   0.4 $  \\
$0.04$  & $   0.76466( 13)$ & $ -13.2 \pm   1.2 $    
        & $ -10.7 \pm   0.6 $ & $  -9.3 \pm   0.3 $  \\
$0.05$  & $   0.70563( 12)$ & $ -16.1 \pm   1.3 $    
        & $ -13.6 \pm   0.6 $ & $ -12.5 \pm   0.4 $  \\
$0.06$  & $   0.65020( 11)$ & $ -18.0 \pm   1.3 $    
        & $ -14.8 \pm   0.7 $ & $ -13.8 \pm   0.4 $  \\
$0.07$  & $   0.59832( 11)$ & $ -18.1 \pm   1.4 $    
        & $ -15.3 \pm   0.7 $ & $ -14.3 \pm   0.4 $  \\
$0.08$  & $   0.54944( 10)$ & $ -18.6 \pm   1.5 $    
        & $ -15.6 \pm   0.8 $ & $ -14.7 \pm   0.5 $  \\
$0.09$  & $   0.50324(  9)$ & $ -18.5 \pm   1.6 $    
        & $ -15.7 \pm   0.8 $ & $ -14.4 \pm   0.5 $  \\
$0.10$  & $   0.45975(  9)$ & $ -19.4 \pm   1.8 $    
        & $ -15.1 \pm   0.9 $ & $ -14.2 \pm   0.6 $  \\
 \hline
 \end{tabular}
 \end{table}

 \begin{table}[htb]
  \caption{The next-to-leading order functions 
  (see eqs.~(\ref{eq:gamma3jets}, \ref{eq:hvaexp}, \ref{eq:Bform})) for
  the DURHAM algorithm.\label{tab:durham}}
 \begin{tabular}{|c|r|r|r|r|}
 \hline
 $y_c$   & $A^{(1)}_V \: \:$ &
 \multicolumn{3}{c|}{$B^{(1)}_V/A^{(1)}_V$}
 \\\cline{3-5}&
 & $M_b=3$GeV & $M_b=4$GeV & $M_b=5$GeV
 \\ \hline
$0.01$  & $    38.229( 56)$ & $  14.7 \pm   2.3 $    
        & $  12.6 \pm   1.2 $ & $   9.6 \pm   0.7 $  \\
$0.02$  & $    34.531( 28)$ & $  -2.2 \pm   1.7 $    
        & $  -2.5 \pm   0.8 $ & $  -3.3 \pm   0.5 $  \\
$0.03$  & $    28.488( 20)$ & $  -5.6 \pm   1.6 $    
        & $  -4.5 \pm   0.8 $ & $  -4.7 \pm   0.5 $  \\
$0.04$  & $    23.521( 16)$ & $  -6.9 \pm   1.7 $    
        & $  -5.2 \pm   0.9 $ & $  -4.8 \pm   0.5 $  \\
$0.05$  & $    19.574( 13)$ & $  -6.5 \pm   1.8 $    
        & $  -4.9 \pm   0.9 $ & $  -4.7 \pm   0.6 $  \\
$0.06$  & $    16.425( 11)$ & $  -5.4 \pm   2.0 $    
        & $  -4.3 \pm   1.0 $ & $  -4.0 \pm   0.6 $  \\
$0.07$  & $    13.878( 10)$ & $  -3.9 \pm   2.1 $    
        & $  -3.6 \pm   1.1 $ & $  -4.0 \pm   0.7 $  \\
$0.08$  & $    11.786(  9)$ & $  -8.4 \pm   2.4 $    
        & $  -4.8 \pm   1.2 $ & $  -4.1 \pm   0.7 $  \\
$0.09$  & $    10.045(  8)$ & $  -3.2 \pm   2.6 $    
        & $  -3.0 \pm   1.4 $ & $  -3.2 \pm   0.8 $  \\
$0.10$  & $     8.585(  8)$ & $  -6.1 \pm   2.8 $    
        & $  -4.1 \pm   1.5 $ & $  -3.3 \pm   0.9 $  \\
 \hline
 $y_c$   & $A^{(1)}_A-A^{(1)}_V$ &
 \multicolumn{3}{c|}{$B^{(1)}_A/A^{(1)}_V$}
 \\\cline{3-5}&
 & $M_b=3$GeV & $M_b=4$GeV & $M_b=5$GeV
 \\ \hline
$0.01$ & $   0.82567( 22)$ & $  18.5 \pm   2.3 $    
        & $  15.8 \pm   1.1 $ & $  12.1 \pm   0.6 $  \\
$0.02$  & $   0.70249( 11)$ & $  -0.8 \pm   1.6 $    
        & $  -1.1 \pm   0.8 $ & $  -2.5 \pm   0.5 $  \\
$0.03$  & $   0.61124(  9)$ & $  -5.5 \pm   1.6 $    
        & $  -3.7 \pm   0.8 $ & $  -4.2 \pm   0.5 $  \\
$0.04$  & $   0.53784(  8)$ & $  -8.1 \pm   1.7 $    
        & $  -5.3 \pm   0.9 $ & $  -5.0 \pm   0.5 $  \\
$0.05$  & $   0.47625(  7)$ & $  -8.3 \pm   1.8 $    
        & $  -5.5 \pm   0.9 $ & $  -5.4 \pm   0.6 $  \\
$0.06$  & $   0.42318(  6)$ & $  -7.7 \pm   2.0 $    
        & $  -4.9 \pm   1.0 $ & $  -4.4 \pm   0.6 $  \\
$0.07$  & $   0.37687(  6)$ & $  -6.8 \pm   2.1 $    
        & $  -4.5 \pm   1.1 $ & $  -4.5 \pm   0.6 $  \\
$0.08$  & $   0.33588(  6)$ & $  -8.4 \pm   2.4 $    
        & $  -5.9 \pm   1.2 $ & $  -3.9 \pm   0.7 $  \\
$0.09$  & $   0.29936(  5)$ & $  -5.5 \pm   2.6 $    
        & $  -4.7 \pm   1.3 $ & $  -3.8 \pm   0.8 $  \\
$0.10$  & $   0.26654(  5)$ & $  -9.3 \pm   2.8 $    
        & $  -4.5 \pm   1.5 $ & $  -3.7 \pm   0.9 $  \\
 \hline
 \end{tabular}
 \end{table}

In tables \ref{tab:e}, \ref{tab:em}, \ref{tab:jade} and \ref{tab:durham}
we present the results for the functions $A_V^{(1)}$ and
$B_{V,A}^{(1)}/A_V^{(1)}$ for the different algorithms and
different values of the pole mass of the $b$-quark. For $A_A^{(1)}$ we have 
presented the difference $A_{A}^{(1)}-A_{V}^{(1)}$, which is entirely due to 
the triangle diagrams V12 in fig.~\ref{fig:loops}. This difference is in 
general small and, as we discuss later, for the observables we are 
interested in, their contribution is suppressed.

Using these functions or their derivatives we will
study two interesting observables that can
be used to extract $m_b(m_Z)$ from data at the $Z$ peak.
The observable proposed some time ago to measure
the bottom-quark mass at the $Z$ resonance was the 
ratio~\cite{Bilenkii:1995ad}
\beq
R^{bd}_3 \equiv \frac{\Gamma^b_{3j}(\yc)/\Gamma^b}
{\Gamma^d_{3j}(\yc)/\Gamma^d}~,
\label{eq:r3bd_def}
\eeq
where $\Gamma^q_{3j}$ and $\Gamma^q$ are the three-jet and the
total decay widths of the $Z$-boson into a quark pair of flavour $q$
in a given jet-clustering algorithm.
More precisely, the measured quantity is~\cite{Abreu:1997ey}
\beq
R^{b\ql}_3 \equiv \frac{\Gamma^b_{3j}(\yc)/\Gamma^b}
{\Gamma^{\ql}_{3j}(\yc)/\Gamma^{\ql}}~,
\label{eq:r3bql_def}
\eeq
where now we normalize on the sum over all light flavours
($\ql=u,d,s$).
To a good approximation both observables are related through
\beq
R_3^{b\ql}=R_3^{bd}+\frac{\alpha_s}{\pi} 
\frac{A^{(1)}_A-A^{(1)}_V}
{A^{(0)}+ \frac{\alpha_s}{\pi} A^{(1)}_V}
\left( \frac{g_{A_b}}{g_{V_b}^2+g_{A_b}^2}
- \frac{\sum_{i=uds} g_{A_i}}{\sum_{i=uds}(g_{V_i}^2+g_{A_i}^2)}
\right)~.
\eeq
The values for $A^{(1)}_A-A^{(1)}_V$ can be taken from the tables.
The extra term is mainly due to the fact that, contrary to $R_3^{bd}$,
even in the $M_b \rightarrow 0$ limit the cancellation of triangle
diagrams~\cite{Hagiwara:1991dx} is not complete in $R_3^{b\ql}$.
The difference is small and rather independent of the
$b$-quark mass. In the DURHAM algorithm and at $y_c=0.02$ it gives a
contribution of +0.002.
In the other three algorithms considered this contribution
is smaller than in DURHAM, since it is suppressed by
a larger leading order $A^{(0)}$ function.
Results for $R_3^{bd}$ in the DURHAM algorithm have already
been presented in~\cite{Rodrigo:1996gw,Rodrigo:1997gy}. Here we
present $R^{b\ql}_3$ for the four clustering algorithms discussed
in this paper.

This observable does not allow for a simultaneous analysis of $\as$ and
$M_b$ because the results for the different values of $\yc$ are
correlated. To be able to make a fit for different values of
$\yc$ we need a differential distribution. One can 
define the following ratio of differential two-jet rates
\begin{equation}
D_2^{b\ql} \equiv 
\frac{[\Gamma_{2j}^b(y_c+\Delta y_c/2)
      -\Gamma_{2j}^b(y_c-\Delta y_c/2)]/\Gamma^b}
     {[\Gamma_{2j}^{\ql}(y_c+\Delta y_c/2)
      -\Gamma_{2j}^{\ql}(y_c-\Delta y_c/2)]/\Gamma^{\ql}}~.
\label{eq:d2bdql_def}
\end{equation}
For numerical results we fix $\Delta y_c = .005$.
The two-jet rate at ${\cal O}(\alpha_s^2)$ is calculated from
the three- and four-jet fractions through the identity \eq{eq:twojets}.
As before, there is a small difference between $D_2^{b\ql}$ and
the $D_2^{bd}$ ratio which is even smaller than 
the difference between $R_3^{b\ql}$ and $R_3^{bd}$.

It is important to note that because the particular normalization we have
used in the definition of these two observables
most of the electroweak corrections cancel. 
Therefore, for our estimates it is accurate enough to consider
tree-level values of $g_V$ and $g_A$. In addition, if they are expressed
in terms of the effective weak mixing angle, most of the weak radiative
corrections are also correctly taken into account~\cite{Akhundov:1999zq}.

From the definitions above, \eq{eq:twojets}, \eq{eq:gamma3jets}, 
the values of the different functions given in the tables and using the
known expression for $\Gamma^b$ 
(for mass effects at order $\as$ see, for 
instance~\cite{Bilenkii:1995ad,Djouadi:1990uk}) we obtain the values of the
two observables as a function of $\yc$ and the quark mass.

We use the following expansion in $\alpha_s$ 
for the two observables considered
\bea
R_3^{b\ql} &=& 1 + \frac{\alpha_s(\mu)}{\pi} \: a_0(y_c) + 
r_b \left(b_0(r_b,y_c) + \frac{\alpha_s(\mu)}{\pi} \: b_1(r_b,y_c)\right)~,
\non \\
D_2^{b\ql} &=& 1 + \frac{\alpha_s(\mu)}{\pi} \: c_0(y_c) + 
r_b \left(d_0(r_b,y_c) + \frac{\alpha_s(\mu)}{\pi} \: d_1(r_b,y_c)\right)~.
\label{eq:pole}
\eea
Note that the exact dependence
on the heavy quark mass is kept in the functions $b_0$, $b_1$, $d_0$
and $d_1$, but for later convenience the leading dependence
on $r_b$ has been factorized out. The $a_0$ and $c_0$ functions 
come exclusively from the triangle diagram and as commented before 
give a small contribution.

Using the known relationship between the perturbative pole mass
and the \msb running mass~\cite{Tarrach:1981up},
\beq
M_b^2 = m_b^2(\mu) \left[1+\frac{2\alpha_s(\mu)}{\pi} 
\left(\frac{4}{3} -\log \frac{m_b^2}{\mu^2} \right)\right]~,
\label{eq:poltorunning}
\eeq
we can re-express the same equations in terms of the running mass
$m_b(\mu)$. Then, keeping only terms of order ${\cal O}(\alpha_s)$
we obtain
\bea
R_3^{b\ql} &=& 1 + \frac{\alpha_s(\mu)}{\pi} \: a_0(y_c) + 
\bar{r}_b(\mu) \left( b_0(\bar{r}_b,y_c) + 
\frac{\alpha_s(\mu)}{\pi} \: \bar{b}_1(\bar{r}_b,y_c,\mu) \right)~, \non \\
D_2^{b\ql} &=& 1 + \frac{\alpha_s(\mu)}{\pi} \: c_0(y_c) + 
\bar{r}_b(\mu) \left( d_0(\bar{r}_b,y_c) + 
\frac{\alpha_s(\mu)}{\pi} \: \bar{d}_1(\bar{r}_b,y_c,\mu) \right)~,
\label{eq:MS}
\eea
where $\bar{r}_b(\mu)=m_b^2(\mu)/m_Z^2$ and
\bea
\bar{b}_1(\bar{r}_b,y_c,\mu) &=& 
b_1(\bar{r}_b,y_c) + 2 b_0(\bar{r}_b,y_c) 
\left(\frac{4}{3} - \log \bar{r}_b + \log \frac{\mu^2}{m_Z^2} \right)~,
\non \\ 
\bar{d}_1(\bar{r}_b,y_c,\mu) &=& 
d_1(\bar{r}_b,y_c) + 2 d_0(\bar{r}_b,y_c) 
\left(\frac{4}{3} - \log \bar{r}_b + \log \frac{\mu^2}{m_Z^2} \right)~.
\label{eq:b1bar}
\eea
The connection between pole and running masses is known up to
order $\as^2$, however, consistency of our pure perturbative 
calculation requires we use only the expression above.
Although at the perturbative level both expressions, \eq{eq:pole} and
\eq{eq:MS}, are equivalent, they give different answers since 
different higher order contributions are neglected.
The spread of the results gives an estimate of the size of 
higher order corrections.

We have performed simple fits to the functions $b_0$ and $d_0$,
describing the leading order behaviour of the $R_3^{b\ql}$ and $D_2^{b\ql}$
observables respectively. Also the $a_0$ and $c_0$ contributions
have been parametrized in terms of simple functions.
The pairs of functions ($b_1$,$d_1$) and ($\bar{b}_1$,$\bar{d}_1$) 
give the NLO heavy quark mass corrections when a description
in terms of the pole mass, \eq{eq:pole},
or in terms of the running mass, \eq{eq:MS},
is used.  Although, by using the relationship in \eq{eq:b1bar}.
one can pass from one set of functions to the other, 
we have also fitted them independently.
A Fortran code containing these fits can be obtained from the 
authors on request. 

\mafigura{13.5cm}{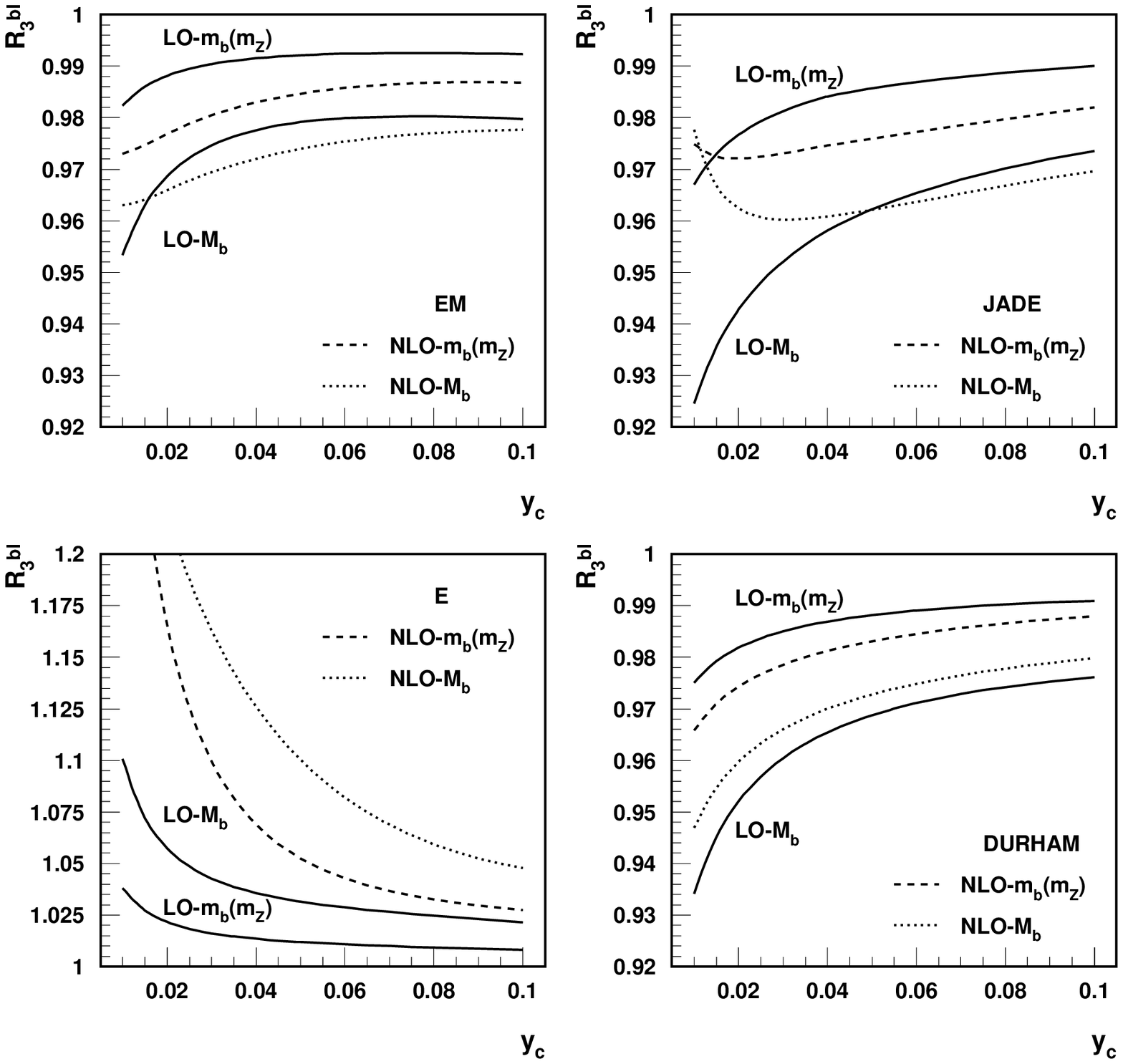}
{The observable $R_3^{b\ql}$ as a function of $\yc$ at the NLO for
the four algorithms studied. The dotted lines give the observable
computed at the NLO in terms of the pole mass $M_b=4.6$~GeV, while
the dashed lines correspond to the use of the 
running mass $m_b(m_Z)=2.83$~GeV instead. In both cases the renormalization 
scale is fixed at $\mu=m_Z$, and $\alpha_s(m_Z)=0.118$. For comparison we also
plot the LO results for $M_b=4.6$~GeV (lower solid lines) and
$m_b(m_Z)=2.83$~GeV (upper solid lines).
}{fig:r3bl}
\mafigura{13.5cm}{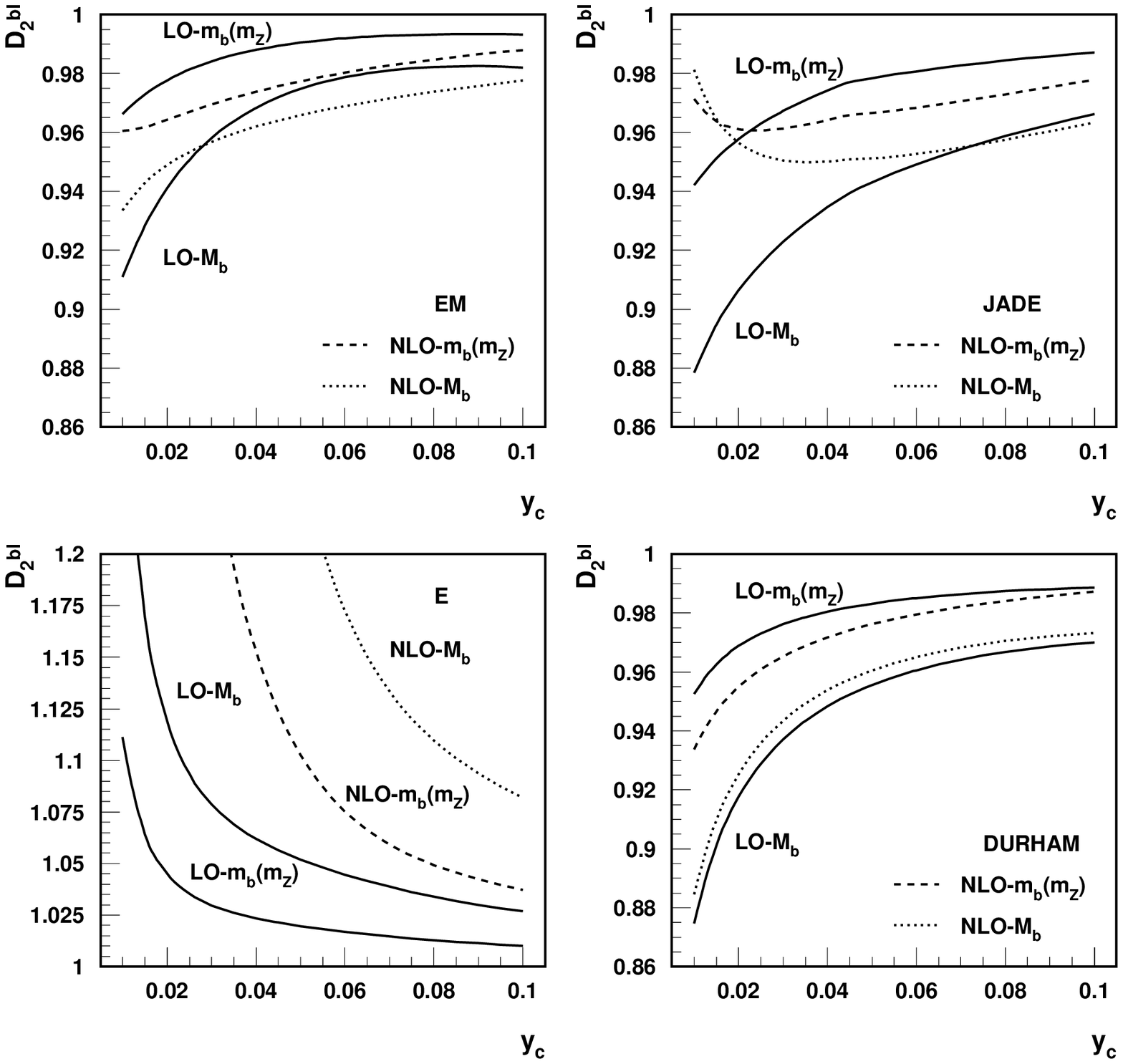}
{Same as in fig.~\ref{fig:r3bl} but for the observable $D_2^{b\ql}$}
{fig:d2bl}

In figs.~\ref{fig:r3bl} and \ref{fig:d2bl} we present the results
for the two observables studied, $R_3^{b\ql}$ and $D_2^{b\ql}$,
in the four jet-clustering algorithms considered.
In all cases we plot the NLO results written either
in terms of the pole mass, 
$M_b=4.6$~GeV~\cite{Jamin:1997rt,Gimenez:1997nw}, or in terms of the running 
quark mass at $m_Z$, $m_b(m_Z)=2.83$~GeV.
The renormalization scale is fixed to $\mu=m_Z$ and
$\alpha_s(m_Z)=0.118$. For comparison we also show $R_3^{b\ql}$
and $D_2^{b\ql}$ at LO when the value of the pole mass, $M_b$, or
the running mass at $m_Z$, $m_b(m_Z)$, is used for the quark mass.

\mafigura{7.cm}{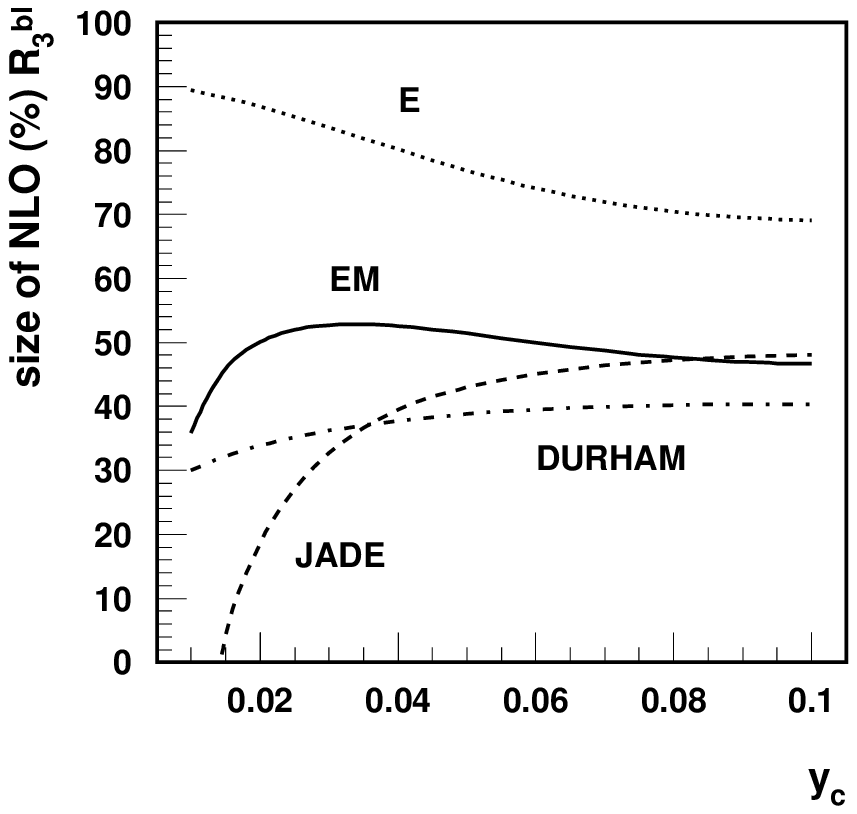}
{Comparative study of the relative size of NLO corrections 
in the four jet algorithms considered, see \eq{eq:size}.}
{fig:size_NLO}

Note the different behaviour of the different algorithms.
In particular the $E$ algorithm.
As already discussed in~\cite{Bilenkii:1995ad},
in this algorithm the shift in the resolution parameter produced
by the quark mass makes the mass corrections positive while 
from kinematical arguments one would expect a negative effect,
since massive quarks radiate less gluons than massless quarks.
Furthermore, the NLO corrections are very large in the $E$ algorithm,
see for instance fig.~\ref{fig:size_NLO}, where we compare  
the following ratio
\beq
\mathrm{size \: of \: NLO} =
\frac{NLO(m_b(m_Z))-LO(m_b(m_Z))}{NLO(m_b(m_Z))-1}~,
\label{eq:size}
\eeq
for $R_3^{b\ql}$ in the four algorithms considered.
The NLO correction in the $E$ algorithm can be as large as
$90~\%$ of the LO result, a fact that probably indicates that 
it is difficult to give an accurate QCD prediction for it.
For the JADE algorithm the NLO correction written in terms of
the pole mass starts to be large for $\yc \le 0.02$.
Note, however that the NLO correction written in terms of the running
mass is still kept in a reasonable range in this region. 
DURHAM, in contrast, is the algorithm that presents a better
behaviour for relatively low values of $\yc$, while keeping
NLO corrections in a reasonable range. 

\mafigura{13.5cm}{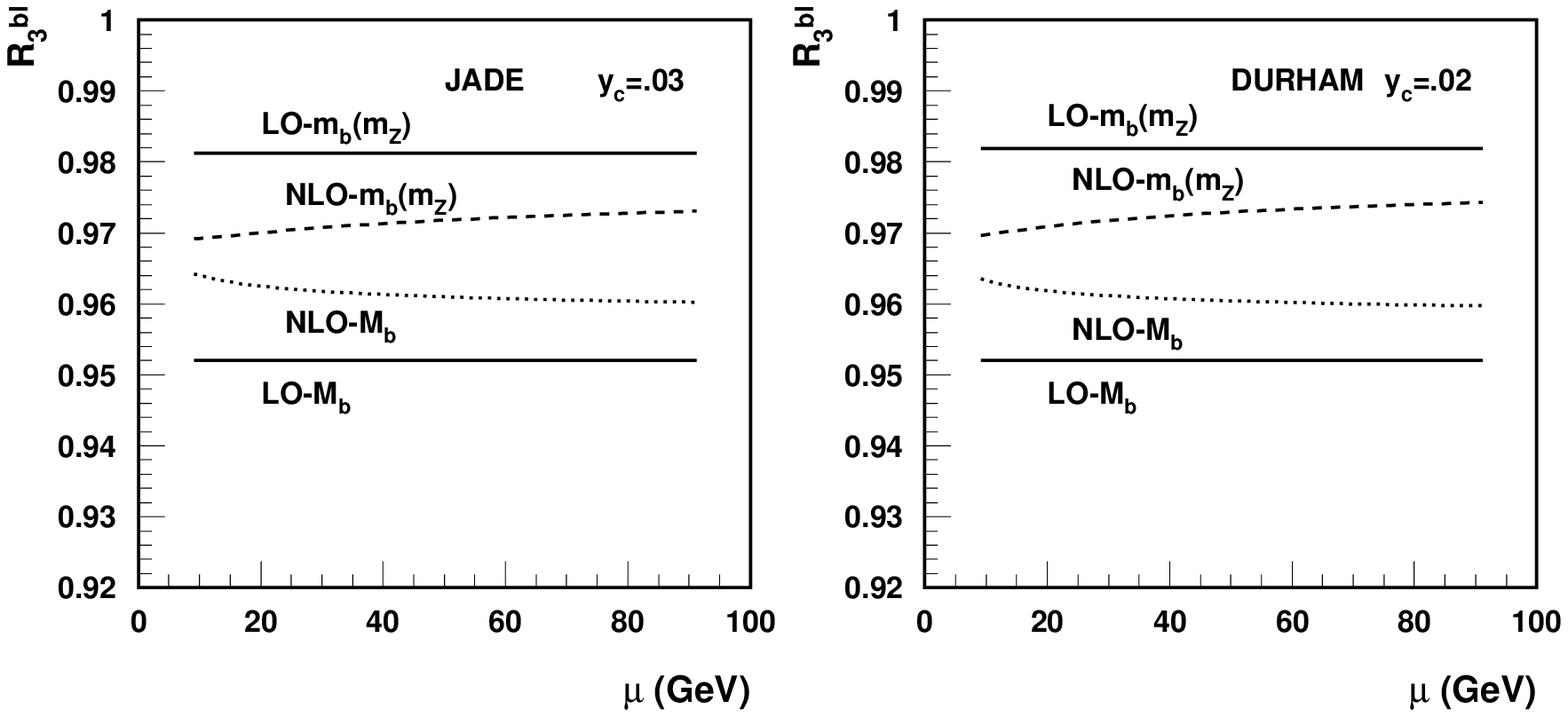}
{The observable $R_3^{b\ql}$ as a function of the renormalization scale
$\mu$ for a fixed value of $\yc$ for the JADE and DURHAM 
algorithms. Labels are as in fig.~\ref{fig:r3bl}.
}{fig:r3bl-mu}

The theoretical prediction for the observables studied
contains a residual dependence on the
renormalization scale $\mu$: when written in terms of the pole mass it only
comes from the $\mu$-dependence in $\alpha_s(\mu)$, when written in terms of
the running mass it comes from both $\alpha_s(\mu)$ and the incomplete
cancellation of the $\mu$-dependences between $m_b(\mu)$ and the logs of $\mu$
which appear in the \msb expression. To give an idea of the uncertainties
introduced by this we plot in fig.~\ref{fig:r3bl-mu}
the observable $R_3^{b\ql}$ as a function of $\mu$ for a fixed value of 
$\yc$. Here we only present plots for the JADE and DURHAM algorithms, which, 
as commented above, have a better behaviour. 
We use the following one-loop evolution equations 
\beq 
a(\mu) = \frac{a(m_Z)}{K}~,
\qquad
m_b(\mu) = m_b(m_Z) \: K^{-\gamma_0/\beta_0}~,
\label{eq:mbrunning}
\eeq
where $a(\mu)=\alpha_s(\mu)/\pi$,
$K = 1 + a(m_Z) \beta_0 \log(\mu^2/m_Z^2)$
with $\beta_0=(11-2/3 N_F)/4$, $\gamma_0=1$ and $N_F=5$ the number of
active flavours, to connect the running parameters at different scales.

\mafigura{13.5cm}{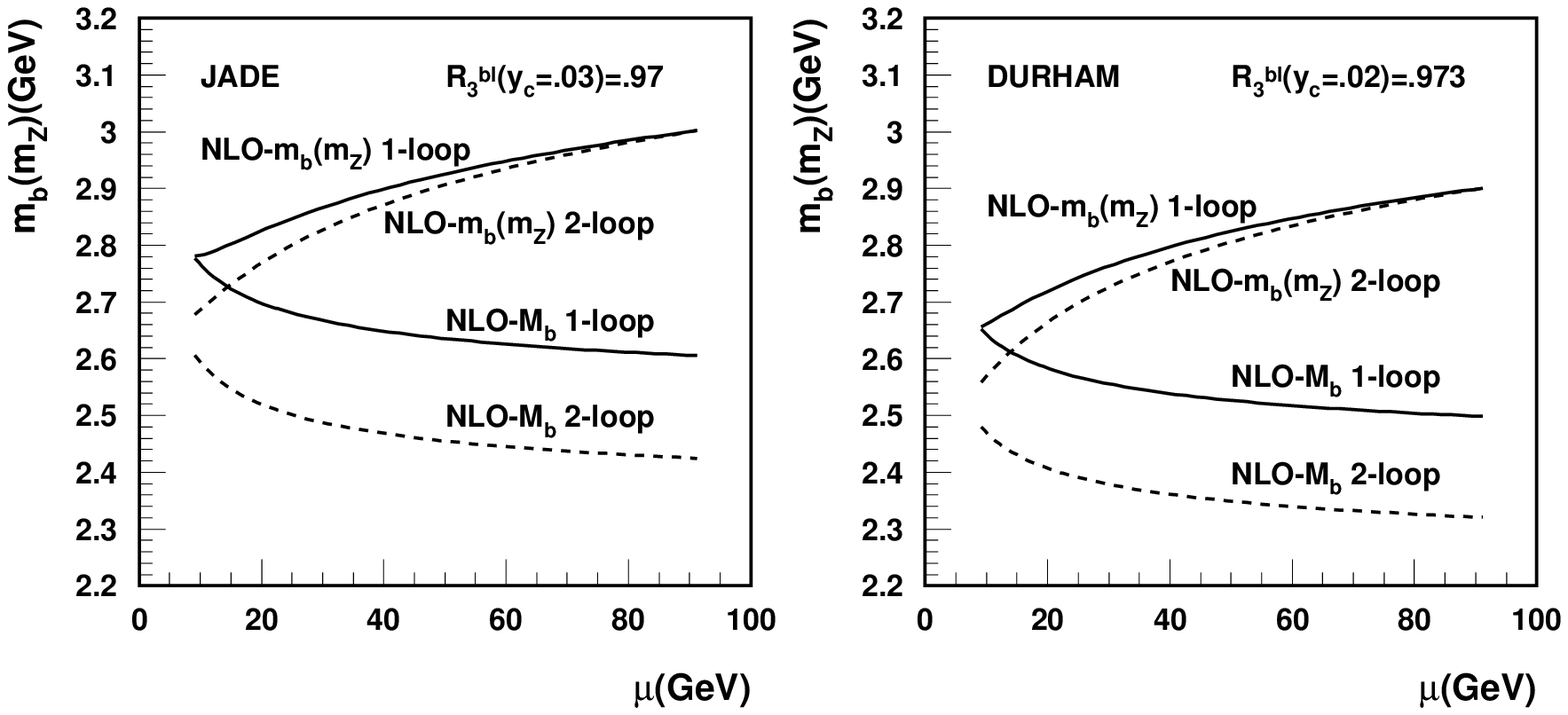}
{Extracted value of $m_b(m_Z)$ from a fixed value of $R_3^{b\ql}$ using
either pole mass expressions (NLO-$M_b$) or running mass expressions
(NLO-$m_b(m_Z)$) as explained in the text. Solid lines are obtained by using
one-loop running to connect the result at different scales. Dashed lines use
two-loop renormalization group equations.}
{fig:mbmz-mu}

For a given value of $R_3^{b\ql}$ we can solve
\eq{eq:pole} (or \eq{eq:MS}) with respect to the quark mass. 
The result, shown in fig.~\ref{fig:mbmz-mu} in the JADE and
DURHAM algorithms for a fixed value of $R_3^{b\ql}$, 
depends on which equation was used and has
a residual dependence on the renormalization scale $\mu$.
The curves in fig.~\ref{fig:mbmz-mu} are obtained in the following way:
first from \eq{eq:MS} we directly obtain for an arbitrary value
of $\mu$ between $m_Z$ and $m_Z/10$ a value for the bottom-quark running
mass at that scale, $m_b(\mu)$, and then using \eq{eq:mbrunning}
we get a value for it at the $Z$-scale, $m_b(m_Z)$.
Second, using \eq{eq:pole} we extract, also for an arbitrary value
of $\mu$ between $m_Z$ and $m_Z/10$, a value for the
pole mass, $M_b$. Then we use \eq{eq:poltorunning} at $\mu=M_b$
and again \eq{eq:mbrunning} to perform the evolution from $\mu=M_b$
to $\mu=m_Z$ and finally get a value for $m_b(m_Z)$.
The two procedures give a different answer since different
higher orders have been neglected in the intermediate steeps. 
The maximum spread of the two results can be interpreted
as an estimate of the size of higher order corrections, that is,
of the theoretical error in the extraction of $m_b(m_Z)$
from the experimental measurement of $R_3^{b\ql}$.
For the taken values of $R_3^{b\ql}$ the spread one gets in $m_b(m_Z)$ 
in both algorithms is roughly a little bit less than $\pm 200$~MeV.
However, the spread of the results is strongly 
dependent on $y_c$ in the JADE algorithm while in DURHAM it 
is almost independent of it. A fact that, once more, shows the good 
theoretical behaviour of the DURHAM algorithm.

Although, our observables are formally order ${\cal O}(\alpha_s)$ 
and therefore compatible with the use of one-loop renormalization 
group equations to connect the running parameters at different 
scales, as a consistency check of the result we have also repeated 
the analysis by using the two-loop running evolution
equations~\cite{Rodrigo:1998zd}
\bea
& & a(\mu) = \frac{a(m_Z)}
{K + a(m_Z) \: b_1 \:  \left( L + a(m_Z) b_1  
\frac{\displaystyle 1-K+L}{\displaystyle K} \right) }~, \non \\
& & m_b(m_Z) = m_b(\mu) \: K^{g_0} \:
\frac{1+a(m_Z) \: c_1}{1+a(\mu) \: c_1}~, 
\eea
where $L=\log K$ and $c_1=g_1-b_1 g_0$ with $b_1=\beta_1/\beta_0$,
$g_i=\gamma_i/\beta_0$ and 
\beq
\beta_1 = \frac{1}{16} \left[102-\frac{38}{3}N_F\right]~, \qquad
\gamma_1 = \frac{1}{16} \left[\frac{202}{3}-\frac{20}{9}N_F\right]~.
\eeq
This corresponds to the dashed lines in Fig~\ref{fig:mbmz-mu}.
In this case the spread of the results is enlarged to
roughly $\pm 280$~MeV mainly due to the change in the NLO-$M_b$
curve. If only the running mass expression, NLO-$m_b(m_Z)$, is used 
the result is more stable when passing from 1-loop to 2-loop 
evolution equations.

\section{Conclusions}

In this paper we presented a detailed description of the
next-to-leading order QCD calculation of the cross-section for heavy
quark three-jet production in $e^+ e^-$ annihilation.
Complete quark mass effects have been taken into account.
As a phenomenological application, we consider
two different observables by using four different jet-clustering
algorithms. The size of unknown higher order corrections is estimated
by studying the residual renormalization
scale dependence and the uncertainty that appears when the
theoretical predictions are written in terms of the running
mass or the pole mass of the produced heavy quark. In particular,
we have carefully studied how to extract a value of the bottom-quark
mass from the experimental measurement of these observables
defined in several jet-clustering algorithms.
These results have already been used in the measurement of the $b$-quark mass 
at the $Z$ peak as well as in the test of the flavour independence of QCD.

\begin{ack}

We are indebted with S. Cabrera, 
J. Fuster and S. Mart\'{\i} for an
enjoyable collaboration. 
This work has been supported by CICYT under the Grant AEN-96-1718, 
by DGESIC under the Grant PB97-1261 and by the Generalitat Valenciana
under the Grant GV98-01-80.
The work of G.R. has also been supported in part by an INFN fellowship.
The work of M.B. was partly supported by the grant GA\v{C}R-2020506.
\end{ack}

\appendix

\section{One-loop integrals contributing to three-parton final states}
\label{ap:loop}

We consider here the basic one-loop integrals 
needed to compute the virtual corrections to the
three-parton decay $q \rightarrow p_1+p_2+p_3$ with all external 
particles on-shell, $p_1^2=p_2^2=M_b^2$ and $p_3^2=0$.
By using the Passarino-Veltman~\cite{Passarino:1979jh} procedure
we reduce any vector or tensor one-loop integral to a combination of
different scalar one-loop integrals.
Furthermore, the one-loop corrected three-parton
amplitudes contribute at second order in the strong coupling constant only
via their interference with the tree-level amplitudes and, therefore, only 
the real parts of scalar one-loop 
integrals~\cite{tHooft:1972fi} are relevant for our calculation.
Since one- and two-point functions are simple 
and well-known we skip their presentation. In this appendix we 
concentrate on the three- and four-point
functions, especially on those with infrared divergences.
Following results for the one-loop functions are understood to refer just to 
their real part. Details of the calculation  
have been presented in~\cite{Rodrigo:1997gv}.

\subsection{Three-point functions}

We define the following set of three-propagator one loop integrals
\bea
\frac{i}{16\pi^2} C01(y_{13}) &=& \mu^{4-D} \int \frac{d^D k}{(2\pi)^D}
           \frac{1}{k^2 [(k+p_{13})^2-M_b^2][(k-p_2)^2-M_b^2]}~, \non
\\ & & \non \\
\frac{i}{16\pi^2} C02(y_{13}) &=& \mu^{4-D} \int \frac{d^D k}{(2\pi)^D}
           \frac{1}{k^2 [(k+p_{13})^2-M_b^2][(k+p_1)^2-M_b^2]}~, \non
\\ & & \non \\
\frac{i}{16\pi^2} C03(y_{13}) &=& \mu^{4-D} \int \frac{d^D k}{(2\pi)^D}
           \frac{1}{k^2 (k-p_3)^2 [(k+p_1)^2-M_b^2]}~,
\\ & & \non \\
\frac{i}{16\pi^2} C04(y_{12}) &=& \mu^{4-D} \int \frac{d^D k}{(2\pi)^D}
           \frac{1}{[k^2-M_b^2][(k+q)^2-M_b^2][(k+p_{12})^2-M_b^2]}~, \non
\\ & & \non \\
\frac{i}{16\pi^2} C05(y_{12}) &=& \mu^{4-D} \int \frac{d^D k}{(2\pi)^D}
           \frac{1}{k^2[(k+p_1)^2-M_b^2][(k-p_2)^2-M_b^2]}~, \non   
\label{eq:allC0}
\eea
where $p_{ij}=p_i+p_j$ and $y_{ij}=2(p_i\cdot p_j)/s$ with $s=q^2$.
In general these functions depend on all the 
masses in the propagators, on the difference of external momenta squared
and on the square of the external 
momenta itself. Since we have fixed masses and we have
imposed the on-shell condition the only remaining 
relevant arguments of these functions are the two-momenta
invariants $y_{13}$ and $y_{12}$, respectively.

For a general scalar three-point function free of infrared
singularities the loop integral can be performed in $D=4$ dimensions
and the result~\cite{tHooft:1972fi} can be expressed
as a sum over twelve dilogarithms (Spence functions).
This is the case for the three-point integrals $C01$, $C02$ and $C04$.
However, after some algebra, the real parts of 
the $C02$ and $C04$ functions take a simpler form
\bea
C02(y_{13}) &=& \frac{1}{s \: y_{13}} \left[ 
\frac{1}{2} \log^2 y_0 + Li_2(y_0) - \frac{\pi^2}{6} \right]~, \non 
\\ & & \non \\
C04(y_{12}) &=& \frac{1}{1-(y_{12}+2r_b)} \frac{1}{2s}
\left[ \log^2 c - \log^2 c_{12} \right]~,
\eea
where
\bea
& & c = \frac{1-\beta}{1+\beta}~, \qquad \qquad
\beta = \sqrt{1-4r_b}~, \non \\
& & c_{12} = \frac{1-\beta_{12}}{1+\beta_{12}}~, \qquad
\beta_{12} = \sqrt{1-\frac{4r_b}{y_{12}+2r_b}}~, \non
\eea
and $y_0 = r_b/(r_b+y_{13})$.
For the $C01$ function we get
\bea
C01(y_{13}) &=& \frac{1}{s \sqrt{\lambda}} \mrm{Re}
\left\{ Li_2\left(\frac{x_1-1}{x_1-z_{11}} \right) 
      - Li_2\left(\frac{x_1}{x_1-z_{11}} \right)
\right. \non \\ & &
      + Li_2\left(\frac{x_1-1}{x_1-z_{12}} \right) 
      - Li_2\left(\frac{x_1}{x_1-z_{12}} \right)
\non \\ & &
      - Li_2\left(\frac{x_2-1}{x_2-z_2} \right) 
      + Li_2\left(\frac{x_2}{x_2-z_2} \right)
\non \\ & & \left.
      - Li_2\left(\frac{x_2-1}{x_2} \right) 
      + 2 Li_2\left(\frac{x_3-1}{x_3} \right) - \frac{\pi^2}{6} \right\}~,
\eea
where 
\bea
x_1 &=& \frac{\alpha-2r_b}{\sqrt{\lambda}}~,    \qquad
x_2 = \frac{\alpha}{1-\alpha} \frac{y_{13}}{\sqrt{\lambda}}~, \qquad
x_3 = -\frac{y_{13}}{\sqrt{\lambda}}~,    \non \\ 
z_{1j} &=& \frac{1}{2}(1 \pm \beta)~, \qquad 
z_2 = \frac{y_{13}}{y_{13}+r_b}~,
\eea
with $\lambda = (1-y_{13})^2-4 r_b$ and $\alpha=(1-y_{13}+\sqrt{\lambda})/2$. 

For the $C03$ function, which has a double infrared pole,
the result we obtain for the real part, after integration
in $D=4-2\epsilon$ dimensions, is
\bea
C03(y_{13}) &=& s^{-1-\epsilon} \mu^{2\epsilon}
\frac{(4\pi)^{\epsilon}}{\Gamma(1-\epsilon)} 
\frac{1}{y_{13}} \\
&\times & \left[ \frac{1}{2\epsilon^2} 
+ \frac{1}{2\epsilon} \log \frac{r_b}{y_{13}^2}
+ \frac{1}{4} \log^2 \frac{r_b}{y_{13}^2} \right.
- \left. \frac{1}{2} \log^2 y_0
- Li_2(y_0) - \frac{7\pi^2}{12} \right]~. \non
\label{eq:C03}
\eea
Notice that for convenience we have factorized a 
$(4\pi)^{\epsilon}/\Gamma(1-\epsilon)$ 
coefficient because this constant appears also in the four-body
phase-space, see \eq{eq:PSsystem34}.

The $C05$ function has a simple infrared pole and already appeared in the NLO
calculation of two-jet production, although in a different kinematical regime.
Here we have
\bea
C05(y_{12}) &=& s^{-1-\epsilon} \mu^{2\epsilon}
\frac{(4\pi)^{\epsilon}}{\Gamma(1-\epsilon)}
\frac{1}{(y_{12}+2r_b)\beta_{12}} \non \\ 
&\times& \left[ \left( 
\frac{1}{\epsilon} - \log r_b \right) \log c_{12}
- 2 L(y_{12}) \right]~,
\label{eq:C05}
\eea
where
\beq
L(y_{12}) = Li_2(c_{12}) + \frac{\pi^2}{3} 
+ \log (1-c_{12}) \log c_{12} - \frac{1}{4} \log^2 c_{12}~.
\eeq

\subsection{Four-point functions}

Only two types of box integrals appear in
the virtual corrections to the three-parton decay. 
We define  
\bea
\frac{i}{16\pi^2} & & D05(y_{13},y_{12}) = \\ & & 
\mu^{4-D} \int \frac{d^D k}{(2\pi)^D} 
\frac{1}{k^2 [(k+p_1)^2-M_b^2][(k+p_{13})^2-M_b^2][(k-p_2)^2-M_b^2]}~, \non
\eea
and
\bea
\frac{i}{16\pi^2} & & D06(y_{13},y_{23}) = \\ & &  
\mu^{4-D} \int \frac{d^D k}{(2\pi)^D} 
\frac{1}{k^2 (k+p_3)^2 [(k+p_{13})^2-M_b^2][(k-p_2)^2-M_b^2]}~, \non
\eea
where $p_{13}=p_1+p_3$ and $y_{ij}=2(p_i\cdot p_j)/s$ with $s=q^2$.

Both four-point functions are IR divergent. The integral $D05$ has
a simple infrared pole, while $D06$, that involves a three-gluon vertex,
exhibits a double infrared pole.
Furthermore, their divergent behaviour is closely related to 
the infrared structure of the $C03$ and $C05$ functions defined 
before in a very simple way~\cite{Bern:1993em,Bern:1994kr} 
\bea
s\cdot D05(y_{13},y_{12}) &=& \frac{1}{y_{13}} C05(y_{12}) 
+ \mrm{finite terms}~, \non \\
s\cdot D06(y_{13},y_{23}) &=& \frac{1}{y_{13}} C03(y_{23}) 
+ \frac{1}{y_{23}} C03(y_{13}) 
+ \mrm{finite terms}~.
\eea

The box integral, $D05$, is also met in one-loop electroweak calculations
and it has been calculated~\cite{Beenakker:1990jr}
by using a photon mass infrared regulator. We just quote the result
in dimensional regularization
\bea
D05(y_{13},y_{12}) &=& s^{-2-\epsilon} \mu^{2\epsilon} 
\frac{(4\pi)^{\epsilon}}{\Gamma(1-\epsilon)}
\frac{1}{(y_{12}+2r_b)\beta_{12} y_{13}} \non \\ 
&\times& \bigg[ \left( 
\frac{1}{\epsilon} + \log \frac{r_b}{y_{13}^2} \right) \log c_{12} 
- \pi^2 - \log^2 c + Li_2(1-c_{12}^2) \non \\
& & - 2 Li_2(1-c_{12}~c) - 2 Li_2(1-c_{12}/c)
\bigg]~,
\label{eq:D05}
\eea

Finally, we have the following result for the real part of the
$D06$ function~\cite{Rodrigo:1997gv} 
\bea
D06(y_{13},y_{23}) &=& s^{-2-\epsilon} \mu^{2\epsilon} 
\frac{(4\pi)^{\epsilon}}{\Gamma(1-\epsilon)} 
\frac{1}{y_{13}y_{23}}  \\
&\times & \left[ \frac{1}{\epsilon^2} 
+ \frac{1}{\epsilon} \log \frac{r_b}{y_{13} y_{23}} \right. 
- \left. \log^2 c 
+ \frac{1}{2} \log \frac{r_b}{y_{13}^2} \log \frac{r_b}{y_{23}^2}
-\frac{3\pi^2}{2} \right]~. \non 
\eea

\section{Phase-space in $D=4-2\epsilon$ dimensions}
\label{ap:ps}

The phase-space for $n$-particles in the final state
in arbitrary space-time dimensions, $D$,
has the following general form~\cite{Marciano:1975tv,Gastmans:1976sr}
\bea
dPS(n) &=& (2\pi)^D \mu^{D-4} \prod_{i=1,n} 
\frac{\mu^{4-D} d^{D-1}p_i}{(2\pi)^{D-1}2E_i}
\delta^D \left( q-\sum_{i=1,n}p_i \right)  \\
&=&  (2\pi)^D \mu^{D-4} \prod_{i=1,n}
\frac{\mu^{4-D} d^{D}p_i}{(2\pi)^{D-1}} 
\delta(p_i^2-m_i^2)\Theta(E_i) \delta^D 
\left( q-\sum_{i=1,n}p_i \right)~. \non
\label{eq:phased}
\eea

Let's consider the decay into three particles,
$q \rightarrow p_1 + p_2 + p_3$, where particles 1 and 2 
share the same mass, $p_1^2 = p_2^2 = M_b^2$, and particle 3
is massless, $p_3^2 = 0$. In terms of the two-momenta
invariants $y_{13}=2(p_1 \cdot p_3)/s$ and 
$y_{23}=2(p_2 \cdot p_3)/s$, with $s=q^2$,
we get $(D=4-2\epsilon)$ 
\beq
PS(3) = \frac{s^{1-2\epsilon} \mu^{4\epsilon}}{16(2\pi)^3}
\frac{(4\pi)^{2\epsilon}}{\Gamma(2-2\epsilon)}
\int \theta(h_p) h_p^{-\epsilon} dy_{13} dy_{23}~,
\eeq
where the function $h_p$, which defines the phase-space boundary,
has the form
\beq
h_p = y_{13} y_{23} (1-y_{13}-y_{23}) - r_b (y_{13}+y_{23})^2~.
\label{eq:bornPS}
\eeq

For the case of the decay into two massive and two 
massless particles, $q \rightarrow p_1 + p_2 + p_3 + p_4$
with $p_1^2 = p_2^2 = M_b^2$ and $p_3^2 = p_4^2 = 0$,
and when the two massless particles can become collinear,
it is convenient to write the four-body phase-space as a 
quasi three-body decay
\bea
\qquad \qquad q \rightarrow & & p_{34} + p_1 + p_2    \non \\
                            & & \hookrightarrow p_3 + p_4~. \non
\eea
In the c.m. frame of particles 3 and 4 the four-momenta can
be written as 
\bea
p_1 &=& (E_1, \ldots, 0, {\bf p}_1)~, \non \\ 
p_2 &=& (E_2, \ldots, {\bf p}_2 \sin \beta, {\bf p}_2 \cos \beta)~, 
\non \\
p_3 &=& E_3 (1, \ldots, \sin \theta \cos \theta', \cos \theta)~,  \non \\
p_4 &=& E_4 (1, \ldots, - \sin \theta \cos \theta', - \cos \theta)~, \non 
\eea
where the dots in $p_3$ and $p_4$ indicate $D-3$ unspecified,
equal and opposite angles (in $D$ dimensions) and $D-3$ zeros
in $p_1$ and $p_2$. We will refer to this as the
``3-4 system''~\cite{Ellis:1981wv}.

In terms of the following invariants
\beq
y_{34}  = \frac{2 (p_3 \cdot p_4)}{s}~,    \qquad 
y_{134} = \frac{2 (p_1 \cdot p_{34})}{s}~, \qquad 
y_{234} = \frac{2 (p_2 \cdot p_{34})}{s}~,
\eeq
where $p_{34}=p_3+p_4$, energies and three-momenta become in the 
``3-4 system''
\bea
E_1 &=& \frac{y_{134}\sqrt{s}}{2 \sqrt{y_{34}}}~, \qquad
{\bf p}_1 = 
\frac{\sqrt{s}}{2 \sqrt{y_{34}}} \sqrt{y_{134}^2-4r_b y_{34}}~, \non \\
E_2 &=& \frac{y_{234}\sqrt{s}}{2 \sqrt{y_{34}}}~, \qquad
{\bf p}_2  = \frac{\sqrt{s}}{2 \sqrt{y_{34}}} \sqrt{y_{234}^2-4r_b y_{34}}~,
\non \\
E_3 &=& E_4 = \frac{\sqrt{y_{34}\, s}}{2}~. \non
\eea
Defining $v=(1-\cos \theta)/2$, the $D$-dimensional 
phase-space in this system is
\bea
PS(4) &=& \frac{s^{1-2\epsilon} \mu^{4\epsilon}}{16(2\pi)^3}
\frac{(4\pi)^{2\epsilon}}{\Gamma(2-2\epsilon)}
\int dy_{134} dy_{234}
\non \\
&\times & \: \frac{1}{2} \: \frac{s^{1-\epsilon} \mu^{2\epsilon}}{16 \pi^2}  
\frac{(4\pi)^\epsilon}{\Gamma(1-\epsilon)}
\int dy_{34} \theta(h_{34}) h_{34}^{-\epsilon} y_{34}^{-\epsilon} 
\non \\
&\times & \int_0^1 dv (v(1-v))^{-\epsilon}
\frac{1}{N_{\theta'}} \int_0^{\pi} d\theta' \sin^{-2\epsilon} \theta' ~,
\label{eq:PSsystem34}
\eea
where a $1/2$ statistical factor has been included, 
$N_{\theta'}$ is a
normalization factor
\beq
N_{\theta'} = \int_0^{\pi} d\theta' \sin^{-2\epsilon} \theta' 
= 2^{2\epsilon} \pi 
\frac{\Gamma(1-2\epsilon)}{\Gamma^2(1-\epsilon)}~,
\eeq
and the function
\bea
h_{34} &=& y_{134}y_{234}(1-y_{134}-y_{234})-r_b(y_{134}+y_{234})^2 \non \\
&-& \left( 1-4r_b-2(1-2r_b)(y_{134}+y_{234})
+ y_{134}^2+3y_{134}y_{234}+y_{234}^2 \right) y_{34} \non \\
&+& 2 (1-2r_b-y_{134}-y_{234}) y_{34}^2-y_{34}^3~,
\label{eq:h34}
\eea
defines the limits of the phase-space.

For the integration of the parts of the four-parton transition probability
containing only quark-gluon soft singularities 
it is convenient to use another parameterization of
the phase-space. For example, if the invariant $y_{13}=2(p_1 p_3)/s$ 
approaches zero, we choose the so-called ``1-3 system''~\cite{Ellis:1981wv}.
Introducing variables
\beq
y_{13}  = \frac{2 (p_1 \cdot p_3)}{s}~,    \qquad
y_{213} = \frac{2 (p_2 \cdot p_{13})}{s}~, \qquad
y_{413} = \frac{2 (p_4 \cdot p_{13})}{s}~, 
\eeq
with $p_{13}=p_1+p_3$.  Energies and three-momenta in this system read 
\bea
E_1 &=& \frac{(y_{13}+2r_b)\sqrt{s}}{2 \sqrt{y_{13}+r_b}}~, \qquad 
\mrm{\bf p}_1  = \frac{y_{13}\sqrt{s}}{2 \sqrt{y_{13}+r_b}}~, \non \\
E_2 &=& \frac{y_{213}\sqrt{s}}{2 \sqrt{y_{13}+r_b}}~,  \qquad 
\mrm{\bf p}_2  = 
\frac{\sqrt{s}}{2 \sqrt{y_{13}+r_b}} \sqrt{y_{213}^2-4r_b(y_{13}+r_b)}~,
\non \\
E_3 &=& \frac{y_{13}\sqrt{s}}{2 \sqrt{y_{13}+r_b}}~,  \qquad 
E_4  = \frac{y_{413}\sqrt{s}}{2 \sqrt{y_{13}+r_b}}~. \non
\eea
In this system, the four-body phase-space has the form
\bea
PS(4) &=& \frac{s^{1-2\epsilon} \mu^{4\epsilon}}{16(2\pi)^3}
\frac{(4\pi)^{2\epsilon}}{\Gamma(2-2\epsilon)}
\int dy_{213} dy_{413} 
\non \\
&\times & \: \frac{1}{2} \: \frac{s^{1-\epsilon} \mu^{2\epsilon}}{16 \pi^2}  
\frac{(4\pi)^\epsilon}{\Gamma(1-\epsilon)}
\int dy_{13} \theta(h_{13}) h_{13}^{-\epsilon}  
\frac{y_{13}^{1-2\epsilon}}{(y_{13}+r_b)^{1-\epsilon}} 
\non \\
&\times & \int_0^1 dv (v(1-v))^{-\epsilon}
\frac{1}{N_{\theta'}} \int_0^{\pi} d\theta' \sin^{-2\epsilon} \theta' ~, 
\eea
where now the angle $\theta$ is the one defined by 
the three-momenta $p_1$ and $p_2$, and the function that 
defines the limits of the phase-space is
\bea
h_{13} &=& (y_{213}y_{413}+4r_b^2) (1-y_{213}-y_{413}) -4r_b^3 \non \\
&-& r_b \left( 1-2(y_{213}+y_{413})+y_{213}^2+4y_{213}y_{413}+2y_{413}^2 
\right) \non \\
&-& \left(1-6r_b+8r_b^2 - 2(1-3r_b)(y_{213}+y_{413})+y_{213}^2
+3y_{213}y_{413}+y_{413}^2 \right) y_{13} \non \\
&+&\left(2-5r_b-2y_{213}-2y_{413} \right) y_{13}^2 - y_{13}^3~.
\eea

\section{Infrared divergent phase-space integrals}
\label{ap:divs}

We consider now the tree-level four-parton decay
$q \rightarrow p_1+p_2+p_3+p_4$.
When at least one gluon is radiated from an external quark (or gluon) line,
the propagator-factors, $1/(p_i\cdot p_j)$, 
can generate soft and collinear divergences at the border 
of the four-body phase-space. In this appendix we present
the basic results that appear when these factors are integrated 
in a thin slice at the border of the phase-space and we 
show how these basic phase-space integrals are related to the
infrared divergent scalar one-loop integrals discussed in
appendix~\ref{ap:loop}.

\subsection{Integrals containing soft gluon divergences}

Phase-space of $n+1$ particles can be written as the product of  
the $n$-body phase-space times the integral over the 
energy and the solid angle of the extra particle.
In arbitrary $D=4-2\epsilon$ dimensions we have
\beq
dPS(n+1) = \frac{\mu^{2\epsilon}}{2(2\pi)^{D-1}}
E^{D-3} \: dE \: d\Omega \: dPS(n)~. 
\eeq
Suppose $E_3$ is the energy of a soft gluon.
Assuming that 
$E_3 < w \sqrt{s}$, where $w$ is very small 
and defines an upper cut on the soft gluon energy.
Then we have the following useful results
\bea
\frac{\mu^{2\epsilon}}{2(2\pi)^{D-1}} & &
\int_0^{w \sqrt{s}}
E_3^{D-3} dE_3 d\Omega \frac{1}{(2 p_1 \cdot p_3)^2} = \non \\
& & \frac{1}{16\pi^2} s^{-\epsilon} \mu^{2\epsilon}
\frac{(4\pi)^{\epsilon}}{\Gamma(1-\epsilon)} 
\frac{1}{E_1^2-{\bf p}_1^2} \non \\
& & \times \left[ - \frac{1}{2\epsilon} + \log 2w
+ \frac{E_1}{2{\bf p}_1} 
\log \frac{E_1-{\bf p}_1}{E_1+{\bf p}_1} + O(w) \right]~,
\label{eq:energy1}
\eea
and
\bea
\frac{\mu^{2\epsilon}}{2(2\pi)^{D-1}} & &
\int_0^{w \sqrt{s}} E_3^{D-3} dE_3 d\Omega 
\frac{1}{(2 p_1 \cdot p_3)(2 p_2 \cdot p_3)} = \non \\
& & \frac{1}{16\pi^2} s^{-1-\epsilon} \mu^{2\epsilon}
\frac{(4\pi)^{\epsilon}}{\Gamma(1-\epsilon)} 
\left[
\left( \frac{1}{2\epsilon} - \log 2w \right) \right. \non \\
& & \times \left. \frac{2}{(y_{12}+2r_b) \beta_{12}} \log c_{12}
+ F(y_{14},y_{24}) + O(w) \right]~. 
\label{eq:energy2}
\eea
Notice that this last integral has the same divergent structure
as the scalar one-loop $C05$ function defined in \eq{eq:C05}.
The finite contribution, the $F(y_{14},y_{24})$ function, is rather
involved. We write it in terms of the variables $y_{13}$ and
$y_{23}$
\beq
F(y_{13},y_{23}) = \frac{1}{b \sqrt{1-a^2}}[G(z_1)+G(z_2)-2G(1)]~,
\eeq
where
\bea
& & G(z) = \log t_1 \log \frac{t_1-z}{t_1+z}
+ \log t_2 \log \frac{z+t_2}{z-t_2} \non \\
&+&  \log \frac{t_1-t_2}{2} 
\log \frac{(t_1+z)(z-t_2)}{(t_1-z)(z+t_2)}  \non \\
&+& \frac{1}{2}
\left[ \log^2 (t_1+z) + \log^2 (t_1-z) \right. \non \\
&+& \left. \log^2 (z-t_2) - \log^2 (z+t_2) \right]  \non \\
&-& \frac{1}{4} \left[ \log (t_1+z) + \log (t_1-z) + 
\log \left( \frac{z-t_2}{z+t_2} \right) \right]^2 \non \\
&+&  Li_2 \left( \frac{z-t_2}{z-t_1} \right)
+ Li_2 \left( \frac{z-t_1}{z+t_1} \right) \non \\
&-& Li_2 \left( \frac{z+t_2}{z+t_1} \right)
- Li_2 \left( \frac{z-t_2}{z+t_2} \right)~,
\eea
with 
\bea 
& & a = \sqrt{h_p}/b~, \\
& & b^2 = 1-4r_b-2(1-2r_b)(y_{13}+y_{23}) \non \\
& & \qquad + (1-r_b) (y_{13}^2+y_{23}^2)
+ y_{13}y_{23}(3-2r_b-y_{13}-y_{23})~, \non 
\eea
where $h_p$ is the function that defines the limits of 
the three-body phase-space, see \eq{eq:bornPS}, and
\bea
& & t_{1,2} = (1 \pm \sqrt{1-a^2})/a~,\\
& & z_1 = \exp \left[ \cosh^{-1} 
\left( \sqrt{(1-y_{13})^2-4r_b}/(a (1-y_{13})) \right) \right]~, \non \\
& & z_2 = \exp \left[ \cosh^{-1} 
\left( \sqrt{(1-y_{23})^2-4r_b}/(a (1-y_{23})) \right) \right]~. \non
\eea

\subsection{Integrals containing gluon-gluon collinear divergences}

For transition probabilities containing gluon-gluon collinear 
divergences we use the four-body phase-space representation 
of \eq{eq:PSsystem34}. In the limit $y_{34}\rightarrow 0$
the function that defines the boundary of the four-body 
phase-space, $h_{34}$ in \eq{eq:h34}, reduces to the 
three-body phase-space function $h_p$ in \eq{eq:bornPS}
and $p_{34}=p_3+p_4$ behaves as the momentum of 
a pseudo on-shell massless particle since $p_{34}^2\rightarrow 0$.
Therefore, in this limit, an effective three-body phase-space 
can be factorized in the same way we made for treating 
the soft singularities.
After integrating over the two angular variables and over 
the $y_{34}$ invariant for $y_{34}<w$ we get the following 
result
\bea
& & dPS(4) \frac{1}{y_{13}y_{34}} =  dPS(3)(y_{134},y_{234})
\: \frac{1}{2} \: \frac{s^{1-\epsilon} \mu^{2\epsilon}}{16 \pi^2} 
\frac{(4\pi)^\epsilon}{\Gamma(1-\epsilon)}
\frac{1}{y_{134}}  \\
& & \times \left[
\frac{1}{2\epsilon^2}-\frac{1}{2} \log^2 w 
+ \left(\frac{1}{2\epsilon}-\log w\right) 
\log \frac{r_b}{y_{134}^2} 
- \frac{1}{4} \log^2 \frac{r_b}{y_{134}^2} 
 - \frac{\pi^2}{4} + O(w) \right]~.  \non
\eea
Notice that we get the same infrared poles as in the one-loop 
three-point function $C03$, \eq{eq:C03}, if we identify 
$y_{134}$ with $y_{13}$.

\bibliographystyle{h-elsevier}
\bibliography{3jetsnlo}

\begin{thebibliography}{10}

\bibitem{Chetyrkin:1996pd}
K.G. Chetyrkin and A. Kwiatkowski,
\newblock Nucl. Phys. B461 (1996) 3, hep-ph/9505358.

\bibitem{Chetyrkin:1997iv}
K.G. Chetyrkin, B.A. Kniehl and M. Steinhauser,
\newblock Phys. Rev. Lett. 79 (1997) 353, hep-ph/9705240.

\bibitem{Chetyrkin:1997vj}
K.G. Chetyrkin and M. Steinhauser,
\newblock Phys. Lett. B408 (1997) 320, hep-ph/9706462.

\bibitem{Rodrigo:1995az}
G. Rodrigo,
\newblock Low-energy {Y}ukawa input parameters for {Y}ukawa coupling
  unification,
\newblock Talk given at International Workshop on Elementary Particle Physics:
  Present and Future, Valencia, Spain, 5-9 Jun 1995. Published in Valencia
  Elem.Part.Phys.1995:0360-370, 1995, hep-ph/9507236.

\bibitem{Langacker:1994xb}
P. Langacker and N. Polonsky,
\newblock Phys. Rev. D49 (1994) 1454, hep-ph/9306205.

\bibitem{Santamaria:1997kd}
A. Santamaria, G. Rodrigo and M. Bilenky,
\newblock Do quark masses run?,
\newblock Talk given at International Workshop on Physics Beyond the Standard
  Model: From Theory to Experiment (Valencia 97), Valencia, Spain, 13-17 Oct
  1997., 1997, hep-ph/9802359.

\bibitem{Bilenkii:1995ad}
M. Bilenky, G. Rodrigo and A. Santamaria,
\newblock Nucl. Phys. B439 (1995) 505, hep-ph/9410258.

\bibitem{Rodrigo:1996gw}
G.V. Rodrigo,
\newblock Quark mass effects in {QCD} jets,
\newblock PhD thesis, Universitat de Val\`encia, 1996, hep-ph/9703359.

\bibitem{MartiiGracia:1997ak}
S. Mart\'{\i}, J. Fuster and S. Cabrera,
\newblock Nucl. Phys. Proc. Suppl. 64 (1998) 376, hep-ex/9708030.

\bibitem{Rodrigo:1997ha}
G. Rodrigo,
\newblock Nucl. Phys. Proc. Suppl. 54A (1997) 60, hep-ph/9609213.

\bibitem{Rodrigo:1997gv}
G. Rodrigo, A. Santamaria and M. Bilenky,
\newblock Dimensionally regularized box and phase space integrals involving
  gluons and massive quarks, 1997, hep-ph/9703360.

\bibitem{Rodrigo:1997gy}
G. Rodrigo, A. Santamaria and M. Bilenky,
\newblock Phys. Rev. Lett. 79 (1997) 193, hep-ph/9703358.

\bibitem{Fuster:1997ah}
J. Fuster, S. Cabrera and S. Marti,
\newblock Nucl. Phys. Proc. Suppl. 54A (1997) 39, hep-ex/9609004.

\bibitem{Abreu:1997ey}
DELPHI, P. Abreu et~al.,
\newblock Phys. Lett. B418 (1998) 430.

\bibitem{Chrin.moriond:1993}
J. Chrin,
\newblock Proc. of the 28th Rencontre de Moriond, Les Arcs, Savoie, France,
  March 1993, edited by J.T.T. Van, p. 313, 1993.

\bibitem{Fuster.jaca:1994}
J. Fuster,
\newblock Recent results on {QCD} at {LEP},
\newblock Proc. XXII International Meeting on Fundamental Physics, Jaca,
  Spain., 1994.

\bibitem{Abreu:1993mc}
DELPHI, P. Abreu et~al.,
\newblock Phys. Lett. B307 (1993) 221.

\bibitem{Buskulic:1995wb}
ALEPH, D. Buskulic et~al.,
\newblock Phys. Lett. B355 (1995) 381.

\bibitem{Beenakker:1989bq}
W. Beenakker et~al.,
\newblock Phys. Rev. D40 (1989) 54.

\bibitem{Laenen:1993zk}
E. Laenen et~al.,
\newblock Nucl. Phys. B392 (1993) 162.

\bibitem{Shifman:1978zq}
M.A. Shifman et~al.,
\newblock Phys. Lett. 77B (1978) 80.

\bibitem{Reinders:1981sy}
L.J. Reinders, S. Yazaki and H.R. Rubinstein,
\newblock Phys. Lett. 103B (1981) 63.

\bibitem{Chang:1982qq}
T.H. Chang, K.J.F. Gaemers and W.L. van Neerven,
\newblock Nucl. Phys. B202 (1982) 407.

\bibitem{Laermann:1980qk}
E. Laermann and P.M. Zerwas,
\newblock Phys. Lett. 89B (1980) 225.

\bibitem{Chetyrkin:1990kr}
K.G. Chetyrkin and J.H. Kuhn,
\newblock Phys. Lett. B248 (1990) 359.

\bibitem{Chetyrkin:1993tx}
K.G. Chetyrkin and A. Kwiatkowski,
\newblock Phys. Lett. B305 (1993) 285.

\bibitem{Ioffe:1978dc}
B.L. Ioffe,
\newblock Phys. Lett. 78B (1978) 277.

\bibitem{Kramer:1980pg}
G. Kramer, G. Schierholz and J. Willrodt,
\newblock Zeit. Phys. C4 (1980) 149.

\bibitem{Rizzo:1980cc}
T.G. Rizzo,
\newblock Phys. Rev. D22 (1980) 2213.

\bibitem{Nilles:1980ic}
H.P. Nilles,
\newblock Phys. Rev. Lett. 45 (1980) 319.

\bibitem{Arbuzov:1992pr}
A.B. Arbuzov, D.Y. Bardin and A. Leike,
\newblock Mod. Phys. Lett. A7 (1992) 2029.

\bibitem{Grunberg:1980ru}
G. Grunberg, Y.J. Ng and S.H.H. Tye,
\newblock Phys. Rev. D21 (1980) 62.

\bibitem{Ballestrero:1992ed}
A. Ballestrero, E. Maina and S. Moretti,
\newblock Phys. Lett. B294 (1992) 425.

\bibitem{Ballestrero:1994dv}
A. Ballestrero, E. Maina and S. Moretti,
\newblock Nucl. Phys. B415 (1994) 265, hep-ph/9212246.

\bibitem{Olsen:1997sk}
H.A. Olsen and J.B. Stav,
\newblock Phys. Rev. D56 (1997) 407.

\bibitem{Bernreuther:1997jn}
W. Bernreuther, A. Brandenburg and P. Uwer,
\newblock Phys. Rev. Lett. 79 (1997) 189, hep-ph/9703305.

\bibitem{Brandenburg:1997pu}
A. Brandenburg and P. Uwer,
\newblock Nucl. Phys. B515 (1998) 279, hep-ph/9708350.

\bibitem{Oleari:1997az}
C. Oleari,
\newblock Next-to-leading order corrections to the production of heavy flavor
  jets in $e^+ e^-$ collisions, 1997, hep-ph/9802431.

\bibitem{Nason:1997nu}
P. Nason and C. Oleari,
\newblock Phys. Lett. B418 (1998) 199, hep-ph/9709358.

\bibitem{Nason:1997nw}
P. Nason and C. Oleari,
\newblock Nucl. Phys. B521 (1998) 237, hep-ph/9709360.

\bibitem{Nason:1997tz}
P. Nason and C. Oleari,
\newblock Phys. Lett. B407 (1997) 57, hep-ph/9705295.

\bibitem{Chrisman:1998}
OPAL, D. Chrisman et~al.,
\newblock QCD98, Montpellier, 1998, 1998.

\bibitem{Abe:1998kr}
SLD, K. Abe et~al.,
\newblock An improved test of the flavor independence of strong interactions,
  1998, hep-ex/9805023.

\bibitem{Burrows:1998kk}
SLD, P.N. Burrows et~al.,
\newblock Heavy quark mass effects and improved tests of the flavor
  independence of strong interactions,
\newblock Contributed paper $\#253$ to the {ICHEP98} Conf., Vancouver,
  (Canada), 1998, hep-ex/9808017.

\bibitem{Giele:1992vf}
W.T. Giele and E.W.N. Glover,
\newblock Phys. Rev. D46 (1992) 1980.

\bibitem{Giele:1993dj}
W.T. Giele, E.W.N. Glover and D.A. Kosower,
\newblock Nucl. Phys. B403 (1993) 633, hep-ph/9302225.

\bibitem{Aversa:1990uv}
F. Aversa et~al.,
\newblock Phys. Rev. Lett. 65 (1990) 401.

\bibitem{Kramer:1989mc}
G. Kramer and B. Lampe,
\newblock Fortschr. Phys. 37 (1989) 161.

\bibitem{Baer:1989jg}
H. Baer, J. Ohnemus and J.F. Owens,
\newblock Phys. Rev. D40 (1989) 2844.

\bibitem{Fabricius:1981sx}
K. Fabricius et~al.,
\newblock Zeit. Phys. C11 (1981) 315.

\bibitem{Gutbrod:1984qa}
F. Gutbrod, G. Kramer and G. Schierholz,
\newblock Z. Phys. C21 (1984) 235.

\bibitem{Bloch:1937pw}
F. Bloch and A. Nordsieck,
\newblock Phys. Rev. 52 (1937) 54.

\bibitem{Kinoshita:1962ur}
T. Kinoshita,
\newblock J. Math. Phys. 3 (1962) 650.

\bibitem{Lee:1964is}
T.D. Lee and M. Nauenberg,
\newblock Phys. Rev. 133 (1964) B1549.

\bibitem{Jersak:1982sp}
J. Jersak, E. Laermann and P.M. Zerwas,
\newblock Phys. Rev. D25 (1982) 1218.

\bibitem{Djouadi:1990uk}
A. Djouadi, J.H. Kuhn and P.M. Zerwas,
\newblock Z. Phys. C46 (1990) 411.

\bibitem{Chetyrkin:1996ia}
K.G. Chetyrkin, J.H. Kuhn and A. Kwiatkowski,
\newblock Phys. Rept. 277 (1996) 189.

\bibitem{Dokshitser:1997in}
Y.L. Dokshitser et~al.,
\newblock JHEP 08 (1997) 001, hep-ph/9707323.

\bibitem{Bentvelsen:1998ug}
S. Bentvelsen and I. Meyer,
\newblock Eur. Phys. J. C4 (1998) 623, hep-ph/9803322.

\bibitem{Moretti:1998qx}
S. Moretti, L. Lonnblad and T. Sjostrand,
\newblock JHEP 08 (1998) 001, hep-ph/9804296.

\bibitem{Bethke:1992wk}
S. Bethke et~al.,
\newblock Nucl. Phys. B370 (1992) 310.

\bibitem{Kunszt:1989km}
Z. Kunszt et~al.,
\newblock {QCD} at {LEP},
\newblock Proceedings of the 1989 LEP Physics Workshop, Geneva, Swizterland,
  Feb 20, 1989, 1989.

\bibitem{Rodrigo:1998nk}
G. Rodrigo, A. Santamaria and M. Bilenky,
\newblock $m_b(m_z)$ from jet production at the {Z} peak in the {C}ambridge
  algorithm, 1998, hep-ph/9807489.

\bibitem{Rodrigo:1998vq}
G. Rodrigo, A. Santamaria and M. Bilenky,
\newblock Improved determination of the b-quark mass at the {Z} peak,
\newblock 4th International Symposium on Radiative Corrections ({RADCOR} 98):
  Applications of Quantum Field Theory to Phenomenology), Barcelona, Catalonia,
  Spain, 8-12 Sep 1998, 1998, hep-ph/9812433.

\bibitem{Bilenky:1998dw}
M. Bilenky, G. Rodrigo and A. Santamaria,
\newblock Heavy quark mass effects in $e^+ e^-$ into three jets,
\newblock To be published in the proceedings of International Euroconference on
  Quantum Chromodynamics ({QCD} 98), Montpellier, France, 2-8 Jul 1998, 1998,
  hep-ph/9811464.

\bibitem{Bilenky:1998dx}
M. Bilenky, G. Rodrigo and A. Santamaria,
\newblock {NLO} calculations of the three jet heavy quark production in $e^+
  e^-$ annihilation: Status and applications,
\newblock Contributed paper $\#988$ to the {ICHEP98} Conf., Vancouver,
  (Canada), 1998, hep-ph/9811465.

\bibitem{Cabrera:1998xx}
DELPHI, S. Cabrera, J. Fuster and S. Mart\'{\i},
\newblock New determination of the $b$-quark mass using improved jet clustering
  algorithms,
\newblock Contributed paper $\#152$ to the {ICHEP98} Conf., Vancouver,
  (Canada), 1998.

\bibitem{Hagiwara:1991dx}
K. Hagiwara, T. Kuruma and Y. Yamada,
\newblock Nucl. Phys. B358 (1991) 80.

\bibitem{Ellis:1981wv}
R.K. Ellis, D.A. Ross and A.E. Terrano,
\newblock Nucl. Phys. B178 (1981) 421.

\bibitem{Berends:1988ab}
F.A. Berends, W.L. van Neerven and G.J.H. Burgers,
\newblock Nucl. Phys. B297 (1988) 429.

\bibitem{tHooft:1972fi}
G. 't~Hooft and M. Veltman,
\newblock Nucl. Phys. B44 (1972) 189.

\bibitem{Marciano:1975tv}
W.J. Marciano and A. Sirlin,
\newblock Nucl. Phys. B88 (1975) 86.

\bibitem{Gastmans:1976sr}
R. Gastmans, J. Verwaest and R. Meuldermans,
\newblock Nucl. Phys. B105 (1976) 454.

\bibitem{Passarino:1979jh}
G. Passarino and M. Veltman,
\newblock Nucl. Phys. B160 (1979) 151.

\bibitem{Chetyrkin:1994js}
K.G. Chetyrkin, J.H. Kuhn and A. Kwiatkowski,
\newblock {QCD} corrections to the $e^+ e^-$ cross-section and the {Z} boson
  decay rate, 1994, hep-ph/9503396.

\bibitem{Stelzer:1994ta}
T. Stelzer and W.F. Long,
\newblock Comput. Phys. Commun. 81 (1994) 357, hep-ph/9401258.

\bibitem{Akhundov:1999zq}
A.A. Akhundov et~al.,
\newblock An effective field theory approach to the electroweak corrections at
  {LEP} energies, 1999, hep-ph/9903546.

\bibitem{Tarrach:1981up}
R. Tarrach,
\newblock Nucl. Phys. B183 (1981) 384.

\bibitem{Jamin:1997rt}
M. Jamin and A. Pich,
\newblock Nucl. Phys. B507 (1997) 334, hep-ph/9702276.

\bibitem{Gimenez:1997nw}
V. Gimenez, G. Martinelli and C.T. Sachrajda,
\newblock Phys. Lett. B393 (1997) 124, hep-lat/9607018.

\bibitem{Rodrigo:1998zd}
G. Rodrigo, A. Pich and A. Santamaria,
\newblock Phys. Lett. B424 (1998) 367, hep-ph/9707474.

\bibitem{Bern:1993em}
Z. Bern, L. Dixon and D.A. Kosower,
\newblock Phys. Lett. B302 (1993) 299, hep-ph/9212308.

\bibitem{Bern:1994kr}
Z. Bern, L. Dixon and D.A. Kosower,
\newblock Nucl. Phys. B412 (1994) 751, hep-ph/9306240.

\bibitem{Beenakker:1990jr}
W. Beenakker and A. Denner,
\newblock Nucl. Phys. B338 (1990) 349.

\end{thebibliography}

\end{document}